\documentclass[12pt]{article}
\usepackage{amsmath, amssymb}
\usepackage{graphicx}
\usepackage{enumerate}
\usepackage{natbib}

\newcommand{\blind}{1}

\usepackage{enumitem}
\usepackage{mathrsfs, booktabs, amsthm, amsfonts, bbold}
\usepackage{color}
\usepackage{hyperref}

\protected\def\[#1\]{\begin{equation}\begin{aligned}#1\end{aligned}\end{equation}}
\protected\def\(#1\){\begin{equation*}\begin{aligned}#1\end{aligned}\end{equation*}}

\usepackage{amsmath}
\usepackage{times}
\usepackage{bm}
\usepackage{natbib}
\usepackage{subcaption}
\usepackage{amssymb}
\usepackage[plain,noend]{algorithm2e}
\usepackage{overpic}

\DeclareMathOperator{\No}{N}

\renewcommand{\t}[1]{\text{#1}}

\usepackage{xcolor}

\newenvironment{psmallmatrix}
  {\left(\begin{smallmatrix}}
  {\end{smallmatrix}\right)}

\newtheorem{theorem}{Theorem}

\theoremstyle{remark}
\newtheorem{remark}{Remark}%

\usepackage{float}

\begin{document}

\def\spacingset#1{\renewcommand{\baselinestretch}%
{#1}\small\normalsize} 
\spacingset{1}

\if1\blind
{
  \title{ Graph-accelerated Markov Chain Monte Carlo using Approximate Samples}
  \author{
	Leo L Duan \thanks{\href{email:li.duan@ufl.edu}{li.duan@ufl.edu}} \\
	Department of Statistics, University of Florida\\and \\
	Anirban Bhattacharya \thanks{\href{email:anirbanb@stat.tamu.edu}{anirbanb@stat.tamu.edu}}
	\\
	Department of Statistics, Texas A\&M University
    }
   \date{}
  \maketitle
  \vspace{-0.5in}
} \fi

\if0\blind
{
  \bigskip
  \bigskip
  \bigskip
  \begin{center}
    {\LARGE\bf Graph-accelerated Markov Chain Monte Carlo \\using Approximate Samples}
\end{center}
} \fi

\bigskip

\begin{abstract}
It has become increasingly easy nowadays to collect approximate posterior samples via fast algorithms such as variational Bayes, but concerns exist about the estimation accuracy. It is tempting to build solutions that exploit approximate samples in a canonical Markov chain Monte Carlo framework. A major barrier is that the approximate sample as a proposal tends to have a low Metropolis-Hastings acceptance rate, as the dimension increases. In this article, we propose a simple solution named graph-accelerated Markov Chain Monte Carlo. We build a graph with each node assigned to an approximate sample, then run Markov chain Monte Carlo with random walks over the graph. We optimize the graph edges to enforce small differences in posterior density/probability between nodes, while encouraging edges to have large distances in the parameter space. The graph allows us to accelerate a canonical Markov transition kernel through mixing with a large-jump Metropolis-Hastings step. The acceleration is easily applicable to existing Markov chain Monte Carlo algorithms. We theoretically quantify the rate of acceptance as dimension increases, and show the effects on improved mixing time. We demonstrate improved mixing performances for challenging problems, such as those involving multiple modes, non-convex density contour, or large-dimension latent variables.\vskip 12pt
\end{abstract}
\noindent
{\it Keywords:} Conductance, Graph Bottleneck, Mixture Transition Kernel, Spanning Tree Graph, Latent Gaussian Model.
\vfill


\newpage

\spacingset{1.9} 

\section{Introduction}
Bayesian approaches are convenient for incorporating prior information, enabling model-based uncertainty quantification, and facilitating flexible model extension. Among the sampling algorithms for posterior computation, Markov chain Monte Carlo is arguably the most popular method and uses a Markov transition kernel (a conditional distribution given the current parameter value) to produce an update to the parameter. As one can flexibly choose transition kernel under a set of fairly straightforward principles, the algorithm can handle parameters in a multi-dimensional and complicated space, such as those involving latent variables \citep{tanner1987calculation,gelman2004parameterization}, constraints \citep{gelfand1992bayesian,duan2020bayesian,presman2023distance}, discrete or hierarchical structure \citep{chib1999mcmc,papaspiliopoulos2007general}; among many others. A major strength is that many Markov chain Monte Carlo algorithms have {an} {\em exact} convergence guarantee \citep{roberts2004general} --- as the number of Markov chain iterations goes to infinity, the {distribution of the} Markov chain samples converge to the target posterior distribution. There is a rich literature on establishing {such} convergence guarantee for popular Markov chain Monte Carlo algorithms. The literature covers Gibbs sampling \citep{roberts1994geometric,gelfand2000gibbs}, Metropolis--Hastings \citep{roberts1994simple,jones2014convergence}, slice sampling \citep{roberts1999convergence,neal2003slice,natarovskii2021geometric}, hybrid Monte Carlo \citep{durmus2017fast,durmus2020irreducibility}, and non-reversible extensions such as piecewise deterministic Markov process \citep{costa2008stability,fearnhead2018piecewise,bierkens2019zig}.

On the other hand, Markov chain Monte Carlo is not without its challenges, especially in advanced models that involve a high-dimensional parameter, latent correlation, or hierarchical structure. For example, there is recent literature characterizing the curse of dimensionality that leads to slow convergence of some routinely used Gibbs sampler \citep{johndrow2019mcmc}, which has subsequently inspired many remedial algorithms \citep{johndrow2020scalable,vono2022efficient}. More broadly speaking, the computing inefficiency happens when the Markov transition kernel creates high auto-correlation along the chain --- under such a scenario, the effective change of parameter becomes quite small over many iterations. Due to this slow mixing issue, practical issues arise in applications: the sampler may take a long time to move away from the start region (where the chain is initialized) into the high posterior density/probability region; it may have difficulty crossing low-probability region that divides multiple posterior modes; it may lack an efficient proposal distribution that could significantly change the parameter value, due to the dependence on a high-dimensional latent variable \citep{rue2009approximate}.

Conventionally, one often views optimization as a competitor to Markov chain Monte Carlo, and as a class of algorithms incompatible with Bayesian models. This belief has been rapidly changed by the recent study of diffusion-based methods. To give a few examples, it has been pointed out that the (unadjusted overdamped) Langevin diffusion algorithm is equivalent to a gradient descent algorithm adding a Gaussian random walk in each step \citep{roberts1996exponential,roberts1998optimal,dalalyan2017theoretical}; as a result, similar acceleration for gradient descent could be applied in the diffusion algorithm for posterior approximation \citep{10.3150/20-BEJ1297}. Mimicking second-order optimization such as Newton descent, one could obtain a rapid diffusion on the probability space using the Fisher information metric \citep{girolami2011riemann}.

In parallel to these developments, variational algorithms have become very popular.
One uses optimization to minimize a statistical divergence between the posterior distribution and a prescribed variational distribution, from which one could draw independent samples as a posterior approximation. The choice of variational distribution spans from mean-field approximation (uncorrelated parameter elements) \citep{blei2017variational}, variational boosting (mixture) \citep{miller2017variational,campbell2019universal}, to normalizing flow neural networks (black-box non-linear transform) \citep{papamakarios2021normalizing}.
In particular, the normalizing flow neural networks have received considerable attention lately due to the high computing efficiency under modern computing platforms, and its high flexibility during density approximation (for continuous parameter).
On the other hand, concerns exist about the accuracy of uncertainty estimates. In particular, the prescribed variational distribution may not be adequately flexible to approximate the target posterior. For example,
the mean-field variational methods lead to a wrong estimate of posterior covariance, which has motivated the development of alternative covariance estimator \citep{giordano2018covariances}. For neural network-based approximation, although positive result has been obtained for approximating the class of sub-Gaussian and log-Lipschitz posterior densities via a feed-forward neural network \citep{lu2020universal}, for normalizing flow (as a neural network restricted for one-to-one mapping), severe limitations have been discovered even for approximating some simple distributions \citep{dupont2019augmented,kong2020expressive}. Practically, another concern is in the lack of diagnostic measures on the accuracy of approximation --- since the target posterior density/probability often contains intractable normalizing constant, usually we do not know how close the minimized statistical divergence is to zero.

Naturally, it is tempting to consider combining strengths from both approximation methods and the canonical Markov chain Monte Carlo framework. One intuitive idea is to adopt the approximate samples to build a proposal distribution and accept or reject each drawn proposal via a Metropolis-Hastings adjustment step. Nevertheless, a technical barrier is the acceptance rate often rapidly decays to zero, as the parameter (and latent variable) dimension grows, which has inspired several solutions.
One remedy is to divide the proposal into several blocks (each in low dimension), then accept or reject the change in each block sequentially via a Metropolis-Hastings-within-Gibbs sampler, which unfortunately often leads to slow mixing. Another idea is to use each approximate sample as an initial state and run multiple parallel Markov chains \citep{hoffman2019neutra}. Lastly, a recently popularized solution is to combine approximate samples with a Metropolis-adjusted diffusion algorithm such as Hamiltonian Monte Carlo \citep{10.3150/16-BEJ810}. For example, \cite{gabrie2022adaptive} interleave two Metropolis-Hastings steps, one using Langevin/Hamiltonian diffusion and one using independent proposal from an approximate sampler (normalizing flow); \cite{Toth2020Hamiltonian} approximate the diffusion by training the gradients of a neural network to match the time derivative of the Hamiltonian, gaining higher efficiency than a differential equation integrator. In addition, there are a few new adaptive Markov chain Monte Carlo algorithms for multi-modal posterior estimation \citep{pompe2020multimode,yi2023global}, based on interleaving the mode estimation steps and proposal moves between modes. The readers can find comprehensive reviews on accelerated Markov chain algorithms in \cite{robert2018accelerating}, and on recent machine learning algorithms in \cite{winter2024emerging}.

Despite similar motivation, our goal is to build a simple and general Markov chain Monte Carlo algorithm for which one could exploit an approximate sampler in an {\em out-of-box manner} without any need for customization. The chosen approximation method  could be as advanced as  a normalizing flow
neural network, or as simple as an existing Markov chain Monte
Carlo (which could suffer from slow mixing). Our main idea is to first collect approximate samples, build a graph to connect these samples and run the Markov chain
Monte Carlo via a mixture transition kernel of a canonical baseline kernel and a graph jump step. We will demonstrate how this method leads to accelerated mixing of the Markov chains in both theory and applications.

\section{Method}

Let $\theta\in \Theta\subseteq \mathbb{R}^p$ be the parameter of interest and our goal is to draw samples from the posterior distribution $\Pi(\theta \mid y)\propto L(y;\theta)\Pi_0(\theta)$, with $L$ the likelihood and $\Pi_0$ the prior. To be general, this form also extends to  augmented likelihood containing latent variable $z$ in addition to the parameter of interest $\tilde \theta$, $\Pi\{ (\tilde\theta,z) \mid y\}\propto L(y,z; \tilde \theta)\Pi_0(\tilde \theta)$, for which one may consider $\theta= (\tilde\theta,z)$. We  use $\Pi$ to represent both distribution and probability kernel (density or mass function). We will primarily focus on continuous $\theta$,  although the method can be extended to discrete $\theta$. 

\subsection{Graph-accelerated Markov Chain Monte Carlo}

Using an existing posterior approximation algorithm {for $\Pi(\cdot \mid y)$}, suppose we have collected $m$ approximate samples $\beta=(\beta^1,\ldots, \beta^m)$. Using those $m$ samples, we first build an undirected and connected graph $G=(V,E_G)$, with {node set} $V=(1,\ldots, m)$, and {edge set} $E_G=\{(i,j)\}${; see Section 2.2 for details}. By connectedness, we mean that for any two  $i$ and $j$, there is a path consisting of edges in $E_G$ between two nodes,  $\t{path}(i,j)= \{ (i,k_1),(k_1,k_2),\ldots, (k_l,j)\}\subseteq E_G$. Accordingly, we define a graph-walk distance between nodes $\t{dist}(i,j)= \min_{\text{all path}(i,j)}|\t{path}(i,j)|$, with $|\cdot|$ the set cardinality, and a ball generated by this distance $B(j;r)=\{k: \t{dist}(j,k) \le r\}$ centered at node $j$ with radius $r$.

We view $(G,\beta)$ as a graph with node attributes; specifically, each $\beta^j$ is a location attribute for node $j$. Taking an existing {\em baseline} Markov chain Monte Carlo algorithm --- such as random-walk Metropolis, Gibbs sampler, or Hamiltonian Monte Carlo sampler, we use $(G,\beta)$ to accelerate the mixing of the Markov chains.
We draw Markov chain samples via a two-component mixture Markov transition kernel:
\[\label{eq:mixture_mtk}
(\theta^{t+1} \mid \theta^t ) \sim \mathcal R(\theta^t, \cdot)= w \mathcal  Q (\theta^{t}, \cdot) + (1-w)\mathcal  K(\theta^{t}, \cdot),
\]
where $w\in [0,1)$ is a tuning parameter. In each iteration, with probability $(1-w)$, the sampler will update $\theta$ using $\mathcal  K(\theta^{t}, \cdot)$, the  Markov chain update step for the baseline algorithm; with probability $w$, the sampler will use $\mathcal Q(\theta^{t}, \cdot)$ to take a {\em graph jump} consisting of the following steps:

\begin{enumerate}
    \item (Project to a node) Find the projection of $\theta^{t}$ to one $\beta^j$, $j =\mathbb{N}(\theta^t):= {\arg\min}_{l} \| \beta^{l} - \theta^{t} \|$.
    \item (Walk on the graph) Draw a new node $i$ uniformly from the ball $B(  j;r)$.
    \item (Relaxation from $\beta^i$) Draw a proposal $\theta^*$ from a relaxation distribution  $F(\theta^* \mid \beta^i,\theta^t)$.
    \item (Metropolis--Hastings adjustment) Accept $\theta^*$ as $\theta^{t+1}$ with probability
    \[\label{eq:detailed_balance}
\alpha(\theta^t,\theta^*)=  \min\bigg [
  1, \frac{ \Pi(\theta^*\mid y)  | B  (i;r ) |^{-1}
        F (\theta^t \mid \beta^{ j},\theta^*)
        }
  { \Pi(\theta^t\mid y)  | B (  j;r )|^{-1}
   F(\theta^* \mid \beta^i ,\theta^t) } \bigg ] 1 \big [\mathbb{N}(\theta^*)=i  \big] ;
  \]
  otherwise keep $\theta^{t+1}$ as the same as $\theta^{t}$.
\end{enumerate}
{Here} $\|a-b\|$ refers to some distance between $a$ and $b$, such as Euclidean distance or {Mahalanobis distance} $\sqrt{(a-b)'S^{-1}(a-b)}$, with $S$ some $p\times p$ positive definite matrix, {for example,} the sample covariance {matrix} based on $\beta$. 
We assume  $\mathbb{N}(\theta)$ is unique almost everywhere with respect to the posterior distribution of $\theta$, and use $F$ to allow $\theta^*$ to take different values from $\beta^i$.  For low-dimensional $\theta$, one could use commonly seen continuous $F$ such as multivariate Gaussian or uniform centered at $\beta^i$. We will discuss specific choices of distance and $F$ suitable for high dimensional $\theta$ in Section 2.3.

\begin{theorem}\label{lemma:detailed_balance}
  The graph jump step satisfies the detailed balance condition: $\Pi(\theta^t\mid y) \mathcal  Q (\theta^{t}, \theta^{t+1}) = \Pi(\theta^{t+1}\mid y) \mathcal  Q (\theta^{t+1}, \theta^{t}).$
\end{theorem}

\begin{remark}
	Since it is possible that the relaxed $\theta^*$ has $\mathbb{N}(\theta^*)\neq i$, we use the indicator function in the acceptance rate to ensure reversibility. An alternative is to use  acceptance rate
	\(
	\min\bigg [
  1, \frac{ \Pi(\theta^*\mid y)  | B [  \mathbb{N}(\theta^*);r ]|^{-1} 
       \sum_{l \in B [  \mathbb{N}(\theta^*);r ]} F (\theta^t \mid \beta^{ l},\theta^*)
        }
  { \Pi(\theta^t\mid y)  | B [  \mathbb{N}(\theta^t);r ]|^{-1} \sum_{i \in B [  \mathbb{N}(\theta);r ]}
   F(\theta^* \mid \beta^i ,\theta^t) } \bigg ],
	\)
	which would be feasible to compute, provided $B [  \mathbb{N}(\theta^t);r ]$ and $B [  \mathbb{N}(\theta^*);r]$ are not too large. For generality, we will use \eqref{eq:detailed_balance} in this article.
	\end{remark}

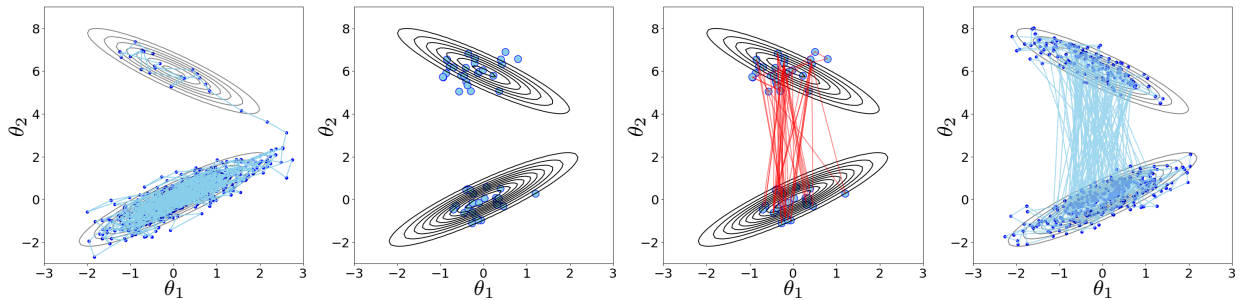
\begin{figure}[H]
 \begin{subfigure}[t]{0.24\textwidth}
 \centering
      \begin{overpic}[width=1.0\linewidth]{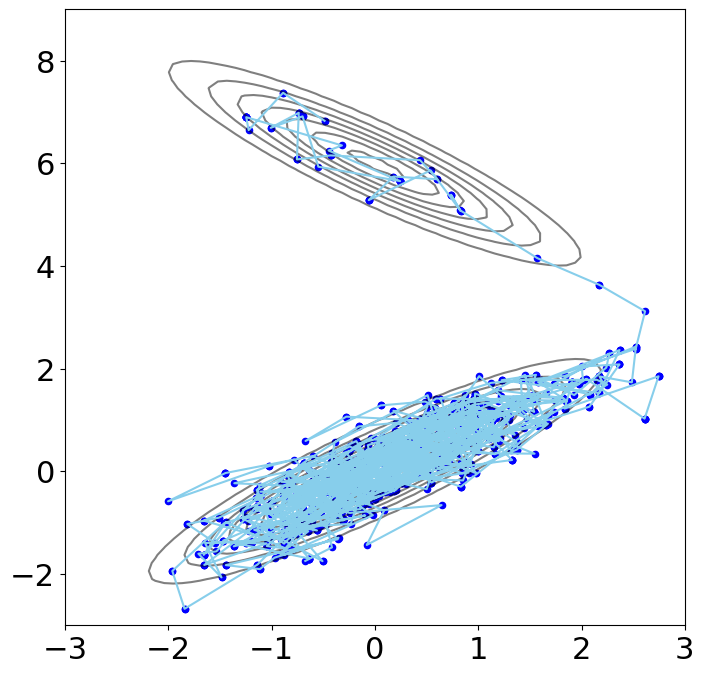}
    \put(50, -4){\scriptsize $\theta_1$}
    \put(-3,50){\rotatebox{90}{\scriptsize $\theta_2$}}
    \end{overpic}
\caption{\footnotesize  Traceplot of a Markov chain produced by a canonical random-walk Metropolis algorithm. The sampler is stuck in one component for a long time, before moving to another.}
 \end{subfigure}\;
   \begin{subfigure}[t]{.24\textwidth}
 \centering
       \begin{overpic}[width=1.0\linewidth]{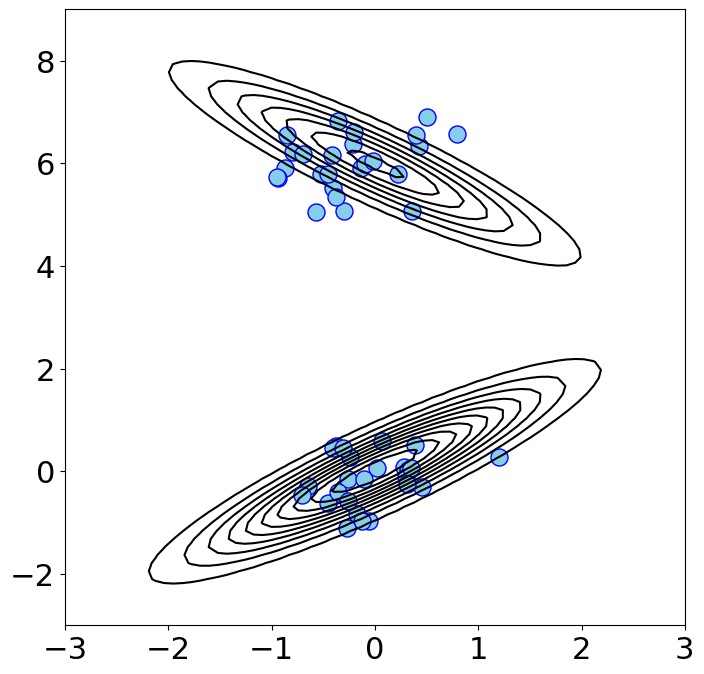}
    \put(50, -4){\scriptsize $\theta_1$}
    \put(-3,50){\rotatebox{90}{\scriptsize $\theta_2$}}
    \end{overpic}
                      \caption{\footnotesize Approximate samples produced by a variational algorithm. Fifty approximate samples are drawn from a two-component Gaussian mixture, with each component having an isotropic covariance and equal mixture weight.}
                       \end{subfigure}\;
 \begin{subfigure}[t]{0.24\textwidth}
 \centering
        \begin{overpic}[width=1.0\linewidth]{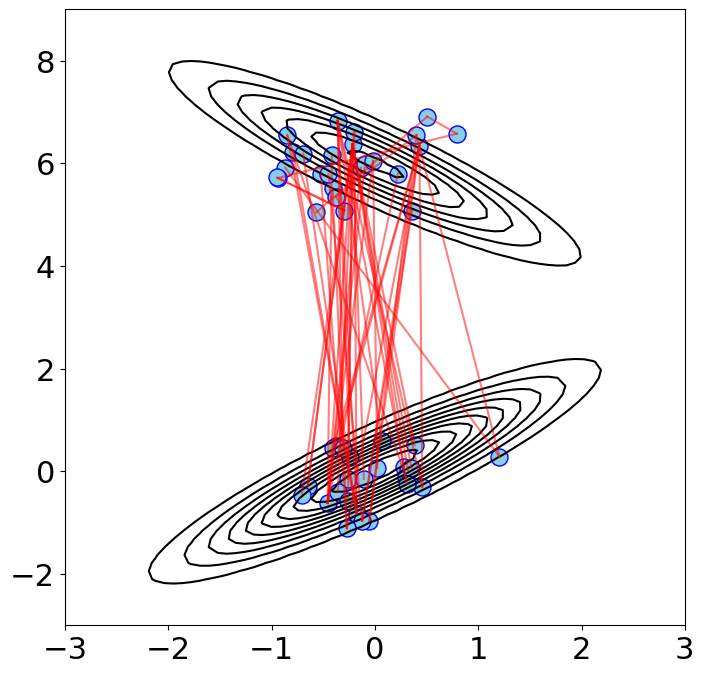}
    \put(50, -4){\scriptsize $\theta_1$}
    \put(-3,50){\rotatebox{90}{\scriptsize $\theta_2$}}
    \end{overpic}
               \caption{\footnotesize A graph connecting the approximate samples, with the graph optimized to increase the parameter distances over the edges, while ensuring small density differences between the parameters on adjacent nodes.}
 \end{subfigure}\;
  \begin{subfigure}[t]{0.24\textwidth}
 \centering
         \begin{overpic}[width=1.0\linewidth]{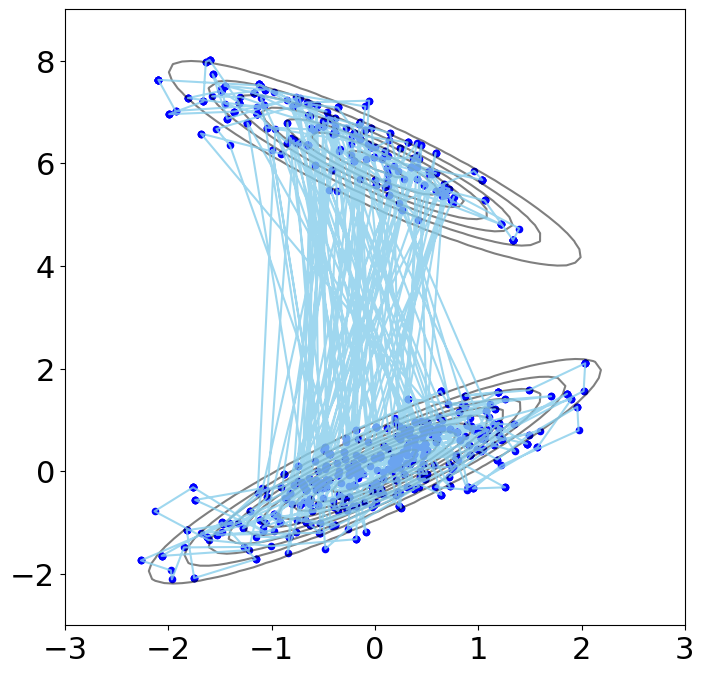}
    \put(50, -4){\scriptsize $\theta_1$}
    \put(-3,50){\rotatebox{90}{\scriptsize $\theta_2$}}
    \end{overpic}
                      \caption{\footnotesize  Traceplot of a Markov chain produced by the graph-accelerated algorithm. The sampler now jumps frequently over the two components.}
 \end{subfigure}
  \caption{An illustrative toy example: using a graph to accelerate the random-walk Metropolis algorithm for Markov chain sampling from a two-component Gaussian mixture distribution. \label{fig:gmm}}
 \end{figure}

To illustrate the idea, we use a toy example of sampling from a two-component Gaussian mixture, $\theta\sim
0.6\t{N}\{  \begin{psmallmatrix}0  \\ 0 \end{psmallmatrix} , \begin{psmallmatrix}1 & 0.9 \\ 0.9 & 1\end{psmallmatrix} \}
+
0.4\t{N}\{  \begin{psmallmatrix}0  \\ 6 \end{psmallmatrix} , \begin{psmallmatrix}1 & -0.9 \\ -0.9 & 1\end{psmallmatrix} \}
$.
 We {consider} the random-walk Metropolis algorithm with  proposal $\theta^*\sim \t{Unif}(\theta^t- \tilde s 1_p, \theta^t + \tilde s 1_p)$, {with the} step size $\tilde s=1$, as the baseline algorithm (corresponding to transition kernel $\mathcal K$). Due to the {high correlation within each mixture component} and the low-density region separating the two modes, the random-walk Metropolis is stuck in one component for a long time as shown in Figure \ref{fig:gmm}(a).

  For acceleration, we use {a} variational distribution $\beta\sim
0.5\t{N}\{  \begin{psmallmatrix}0  \\ 0 \end{psmallmatrix} , \sigma^2_1 I \}
+
0.5\t{N}\{  \begin{psmallmatrix}0  \\ 6 \end{psmallmatrix} , \sigma^2_2 I \}$, with $\sigma^2_1$ and $\sigma_2^2$ numerically calculated via minimizing the Kullbeck-Leibler divergence through the \texttt{numpyro} package \citep{phan2019composable}, and then draw $50$ independent approximate samples {from the resulting variatonal approximation}. We obtain a simple graph $G$, a spanning tree \citep{kruskal1956shortest,prim1957shortest}, that connects all those samples, and use the graph-accelerated algorithm in \eqref{eq:mixture_mtk} with  $w=0.3$ and Gaussian for $F$. As shown in Figure \ref{fig:gmm}(d), the sampler now jumps rapidly over the two components. We run both the baseline and accelerated algorithms for $10,000$ iterations, and using the effective sample size of $\theta_2$ per iteration as a benchmark for mixing: the one of the baseline algorithm is $0.04\%$, and the one of the accelerated version is $4.5\%$, hence is roughly 100 times faster.

 \begin{remark}\label{rm:nonergodicity_of_q}
 	Before we elaborate further on the details, we want to clarify two points to avoid potential confusion. First, since approximate algorithms may produce sub-optimal estimates of the posterior (such as ignoring the covariance as in the above example), we do not want to completely rely on the graph-jump step  $\mathcal Q$ for Markov chain transition. Therefore, we consider a mixture kernel $\mathcal R(\theta^t, \cdot)$ with $w<1$. Second, the graph-jump $\mathcal Q$ itself does not have to lead to an ergodic Markov chain (one that could visit every possible state) --- rather, we use $\mathcal  Q$ to form a network of {\em highways} and allow fast transition from one region to another one far away, while relying on $\mathcal K$ as {\em local roads} for ensure ergodicity.
 \end{remark}

\subsection{Choice of Graph for Fast-mixing Random Walk}\label{subsec:graph}
Given $(\beta^1,\ldots,\beta^m)$, there are multiple ways to form a connected graph $G$. To begin the thought process, one choice for $G$ is the complete graph, in which every pair of nodes is connected by an edge. However, it is not hard to see that a
 $\beta^j$ is likely to have several $\beta^k$'s in $\Theta$-space neighborhood with small $\|\beta^j-\beta^k\|$; intuitively, the values of $\Pi(\beta^k \mid y)$ of those close-by points tend to dominate over the points far away. As a result, a jump over $G$ would likely correspond to a small change and hence be not ideal.

To favor jumps over large  $\|\beta^j-\beta^k\|$ with a simple choice of $G$, we consider the opposite to a complete graph, and focus on the smallest and connected graph: an undirected spanning tree $G$ containing $(m-1)$ edges. The spanning tree enjoys a nice optimization property, that we can easily find the global minimum of a sum-over-edge loss function. As a result, we can customize the loss to balance between the posterior kernel difference and the jump distance. To be concrete, we use the following minimum spanning tree:

\[ & G = \underset{\text{all spanning trees } T}{\arg\min} \sum_{(i,j)\in E_T} c_{i,j}, \\
 & c_{i,j} = \left \{
\begin{array}{ll l}
  {\kappa}/\{ {1+ \|\beta^i -\beta^j\|}\}, & \text{ if } | \log \Pi(\beta^i \mid y) - \log\Pi(\beta^j \mid y)|<\kappa,  \\
    | \log \Pi(\beta^i \mid y) - \log\Pi(\beta^j \mid y)|, & \t{ otherwise}
  \end{array}
  \right.
  \label{eq:mst},
 \]
 with $\kappa>0$ some chosen threshold.
The minimum spanning tree can be found via several algorithms \citep{prim1957shortest,kruskal1956shortest}. We state Prim's algorithm \citep{prim1957shortest} here to help illustrate an insight. One starts with a singleton node set $V_1=\{1\}$ and an empty $E_T$ to initialize the tree, and $V_2=V\setminus V_1$; each time we add a new node $\hat j$ associated with
\(
(\hat i,\hat j)=\underset{(i,j): i\in V_1, j\in V_2}{\arg\min} c_{i,j},
\)
and add it to $E$, and move $\hat j$ from $V_2$ to $V_1$; we repeat until $V_2$ becomes empty. We can see that this algorithm is {\em greedy}, in the sense that it finds the locally optimal $c_{i,j}$ in each step; nevertheless, thanks to the $M$-convexity \citep{murota1998discrete} (a discrete counterpart of continuous convexity) of {the} minimum spanning tree problem, the greedy algorithm will produce a globally optimal tree.

The equivalence between local and global optimality allows us to gain interesting insight --- each time we add a new node to the graph, if there is more than one candidate edge $(i,j): | \log \Pi(\beta^i \mid y) - \log\Pi(\beta^j \mid y)|<\kappa$, then we will choose the one with the largest distance $\|\beta^i-\beta^j\|$. On the other hand, if $\beta^i$ has all $\beta^j:| \log \Pi(\beta^i \mid y) - \log\Pi(\beta^j \mid y)| \ge \kappa$, then we will choose one with the smallest kernel difference. We will quantify the effect of $\kappa$ on acceptance rate in Theorem 1.

In this article, for simplicity, we choose $\kappa=1$ and $r=3$, as they seem adequate to show {impressive} empirical performance. Nevertheless, one may also consider two extensions that could further improve the mixing performance, although the procedures are more complicated. 

First, one could numerically optimize $\kappa>0$ and $r\in \{1,\ldots,m\}$ to approximately maximize the expected squared jumped distance (ESJD), a measure on the mixing of Markov chain \citep{Gelman_Gilks_Roberts_1997,pasarica2010adaptively}. Since at the graph-construction stage, we do not yet have access to Markov chain samples collected from $\mathcal R$, we may use approximate samples $\beta$ to form an empirical estimate for expected squared jumped distance in a random walk on the graph $G_\kappa$  (a graph parameter varying with $\kappa$):
\(
\frac{1}{m}\sum_{j=1}^{m} \frac{1}{B_\kappa(j;r)}\sum_{i\in B_\kappa(j;r)}\min\bigg \{ 1, \frac{\Pi(\beta^i \mid y) | B_\kappa (  i;r )|^{-1}}
{\Pi(\beta^j \mid y) | B_\kappa (  j;r )|^{-1}} \bigg \}
\|\beta^i-\beta^j\|^2,
\)
where we use subscript on $B_{\kappa}(j;r)$, to indicate that the ball varies with the value of $\kappa$. The maximization over $(\kappa,r)$ is non-convex, however, one could obtain local via standard grid search.

\begin{remark}
	If we choose $G$ as a $d$-regular graph (instead of a spanning tree) and $r=1$, we could instead optimize $G$ under degree constraints to directly maximize the empirical ESJD
	\(
\frac{1}{m(d+1)}\sum_{(i,j)\in E_G}\bigg [1+ \min \{ \frac{\Pi(\beta^j \mid y)}{\Pi(\beta^i \mid y)}
, \frac{\Pi(\beta^i \mid y)}
{\Pi(\beta^j \mid y)} \} \bigg ]
\|\beta^i-\beta^j\|^2,
\)
although the optimization of $d$-regular graph is more complex than the one of spanning tree.
\end{remark}

Second, instead of focusing on graph choice, one could generalize and focus on optimizing for a random walk transition probability matrix, equivalently to drawing non-uniform $i\in B(j;r=1)$. To be concrete, consider a given bidirectional and connected graph $\bar G$ of $m$ nodes, we want to estimate a transition probability matrix $P\in [0,1]^{m\times m}$ with $P_{i,j}$ the probability of moving from $i$ to $j$. This matrix satisfies the following constraints:
\(
P1_m=1_m,
 \qquad \pi_\beta^{\rm T} P = \pi^{\rm T}_\beta,
\qquad P_{i,j}=0 \text{ if } (i\to j)\not \in E_{\bar G},
\)
where $\pi_\beta$ is a given target probability vector that we want the random walk to converge to in the marginal distribution ($\pi^{\rm T}_\beta = \lim_{t\to \infty}\pi^{\rm T}_* P^{t}$ for any initial probability vector $\pi^{\rm T}_*$). A sensible specification is $\pi_\beta(j) \propto \Pi(\beta^j \mid y)$. 
The first equality above ensures that $P$ is a valid transition probability matrix, and the second one gives the global balance condition for random walk on $\bar G$.

Since the convergence rate of $\pi^{\rm T}_0 P^{t}$ toward $\pi_\beta$ depends on the second largest magnitude of the eigenvalue of $P$, and its largest eigenvalue $1$ corresponds to right eigenvector $1_m$ and left eigenvector $\pi_\beta$. We can formulate an optimization problem as
\( \hat P= {\arg\min}_P  \| P- 1_m \pi^{\rm T}_\beta \|_2 \)
where $P\in [0,1]^{m\times m}$ is subject to the two constraints above,  $\|.\|_2$ above is the spectral norm. This is a convex problem that can be solved quickly. Note that when $\bar G$ is a complete graph, we would obtain a trivial solution $P=1_m \pi^{\rm T}_\beta$, corresponding to  $P_{i,j}\propto \Pi(\beta^j \mid y)$ for any $i$ --- since under moderate or high dimension, one $\beta^{\hat j}$ will likely dominate over all other $\beta^{j}$'s in posterior density, the trivial solution will likely  always draw node $\hat j$ when forming proposal $\theta^*$, which would not be ideal. Therefore, one may want to exclude from $\bar G$ those $(i\to j)$  corresponding to short distance $\|\beta^i-\beta^j\|$. Once we obtain $\hat P$, we could draw $i$ from $B(j;1)$ with probability $\hat P_{j,i}$ \{replacing $|B(j,r)|^{-1}$ in \eqref{eq:detailed_balance}\}. This extension is inspired by \cite{boyd2004fastest}; nevertheless, the difference is that they focus on the random walk on an undirected graph with $P=P^{\rm T}$, with $\pi_\beta(i) = 1/m$ as the target. We provide the optimization algorithm in the appendix, and numerical illustration in the Supplementary Materials S.1.

\subsection{Choice of Relaxation Distribution for High-dimensional Posterior}
It is known that {Metropolis--Hastings} algorithms, if employed with a fixed step size for the proposal, suffer from the curse of dimensionality: the acceptance rate decays to zero quickly as {dimension} $p$ increases. Based on existing study for Gaussian random-walk Metropolis algorithm with target distribution consisting of $p$ independent components \citep{Gelman_Gilks_Roberts_1997,Roberts_Rosenthal_2001}, we can estimate that the vanishing speed of acceptance rate under a fixed step size is roughly $O\{ \exp(- \tilde c p)\}$ for some constant $\tilde c >0$, with detail given in the Supplementary Materials S.2.

As a result, if we use a continuous $F(\theta^* \mid \beta^i ,\theta^t)$ such as multivariate Gaussian in the graph-jump step,  our algorithm will also suffer from a fast {decay} of acceptance rate as $p$ increase. Therefore, we propose a special relaxation distribution $F$ to slow down the {decay}, as follows.

 \begin{figure}[H]
  \centering
        \includegraphics[width=.5\linewidth]{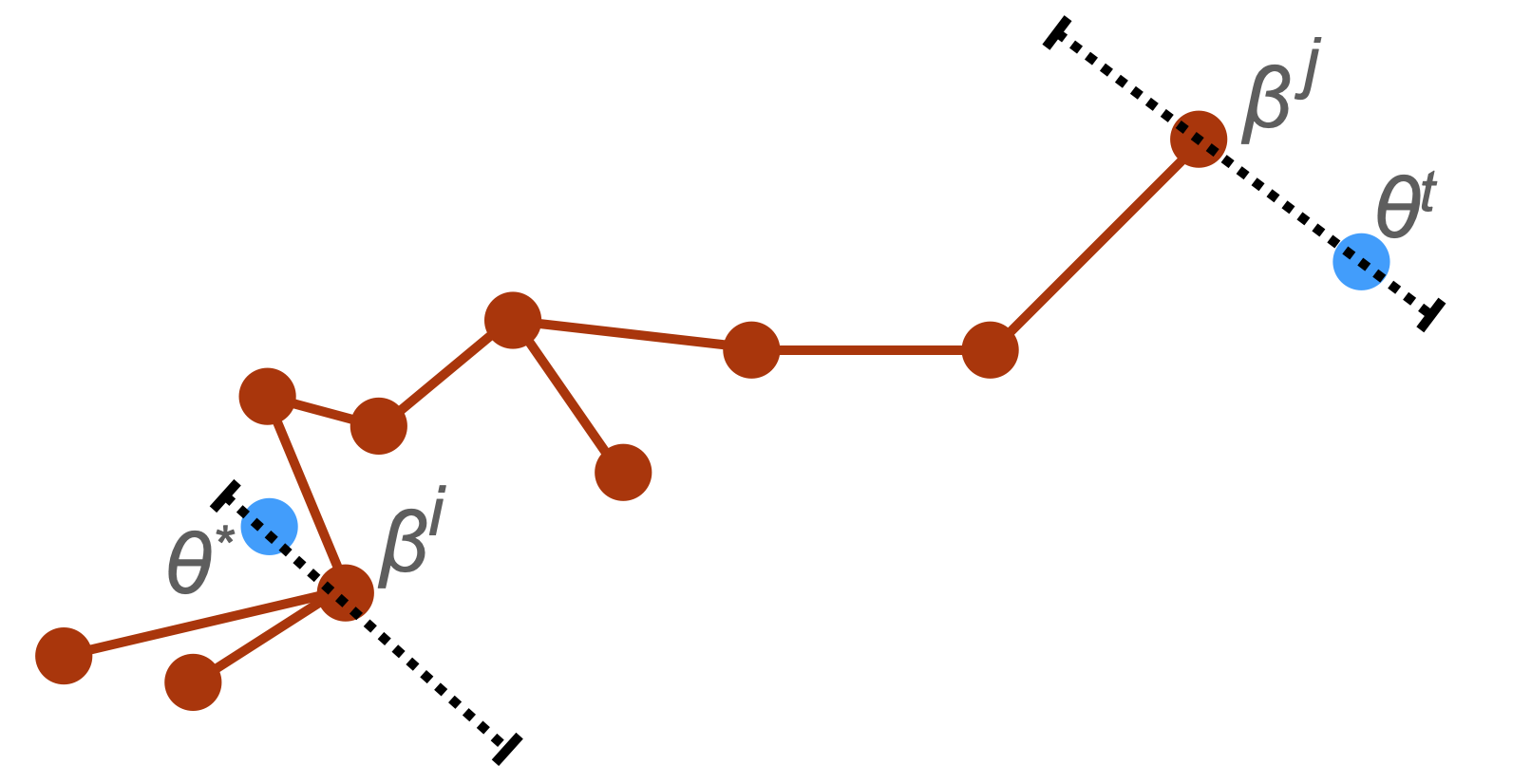}
        \caption{Diagram illustrating the proposal for moving from $\theta^t$ (blue on the right) to $\theta^*$ (blue on the left):
          (i)  project $\theta^t$ to a node on the graph $ j=\mathbb{N}(\theta^t)$ with location attribute  $\beta^{ j}$, find the line crossing  $\theta^t$ and $\beta^{ j}$; (ii) random walk to node $i$, and find the line interval parallel to $(\theta^t-\beta^j)$, containing $\beta^i$ and points $x$ projecting to $\theta^i:$ $\mathbb{N}(x)=i$; (iv) sample $\theta^*$ from the new interval and accept $\theta^{*}$ via Metropolis-Hastings criterion. \label{fig:proposal}}
  \end{figure}
 \begin{itemize}
     \item Find $ j =\mathbb{N}(\theta^t)$, and calculate the directional unit-length vector between $\theta^t$ and its projection $\beta^{ j}$, $v = (\theta^{t} - \beta^{ j})/ \|\theta^{t} - \beta^{ j}\|$.
     \item On the line $\{x: x =  \beta^i + \xi v\}$ with $\xi\in \mathbb{R}$, find the maximum-magnitude $a^i$ and $b^i$, such that
     \(
    \mathbb{N}(\beta^i + \xi v)=i \; \forall \xi\in(a^i,b^i), -l\le a^i\le 0, 0\le b^i\le l.
    \)
     \item Draw $\theta^*$ uniformly from $\{x: x =  \beta^i + \xi v, a^i<\xi<b^i\}$.
 \end{itemize}
  It is not hard to see that 
  \(
 F(\theta^* \mid \beta^i, \theta^t) = \frac{1}{b^i-a^i} 1\{ \|\theta^*-\beta^i\|\le l, \mathbb{N}(\theta^*)= i \},
 \)
 where $l>0$ is a truncation to ensure properness of $F$.  We can find the line segment easily using one-dimensional bisection.  The acceptance rate in \eqref{eq:detailed_balance} becomes:
   \[
   \min\bigg [
  1, \frac{ \Pi(\theta^*\mid y)  | B  (i;r ) |^{-1}
         (b_j-a_j)^{-1}
        }
  { \Pi(\theta^t\mid y)  | B (  j;r )|^{-1}
  (b_i-a_i)^{-1} } \bigg ].
  \]
In general, to deal with the curse of dimensionality problem in Metropolis-Hastings algorithms, one often resorts to the Metropolis-Hastings-within-Gibbs strategy, which divides the parameter into blocks and updates each low-dimensional block sequentially. The strategy can be understood as a way of reducing the dimension of each proposed change from $\theta^t$ to $\theta^*$.

In our relaxation distribution, although the proposed $\theta^*$ is different from $\theta^t$ at almost all of its coordinates, as the directional vector $v$ is completely determined by $\theta^t$, only two univariate random variables are drawn when forming $\theta^*$: the choice of $i$ and shift $\xi\in \mathbb{R}$. Therefore, the effective low dimension gives an intuition about how the specific $F$ slows down the decay of acceptance rate. We will formally quantify the scaling of acceptance rate in the theory section.

\section{Theoretical Results}
In this section, we provide a theoretical exposition of the graph-accelerated algorithm. Compared to the simplicity of the method presented in the previous sections, the results here are more technical and obtained under a few assumptions set up for a tractable mathematical analysis.

\subsection{Accelerated Mixing}
We now focus on the mixing time of the accelerated algorithm. To provide the necessary background, denote the state space by $\Theta$, and consider a Markov transition kernel $\mathcal M$, with $\mathcal M(x,\cdot)$ the transition probability measure from state $x$ and  $\pi(\cdot)$ the invariant distribution of ${\mathcal M}$. Under the context of posterior estimation, we have $\pi(\cdot)$ equal the posterior distribution associated with kernel $\Pi(\theta\mid y)$. We use $\mathcal M^t(x^0, \cdot)$ to denote the distribution after $t$ iterations of transitioning via $\mathcal M$ with $x^0$ an initial point randomly drawn from $\pi^{0}$. Given a small positive number $\eta$, the $\eta$-mixing time of a Markov chain is:
\(
\min \{ t:\sup_{A\in \Theta}| \mathcal M^t(x^0, A) -  \pi(A) | \le \eta\}.
\)
Therefore, at a given $\eta$, the $\eta$-mixing time would be dependent on $\pi^0$ and $\mathcal M$.
Since the left-hand side of the inequality is often intractable, one often derives an upper bound of the left-hand side as a diminishing function of $t$, and produces an upper bound estimate of the mixing time.

Now we review an important concept of {\em conductance}, which is useful for calculating the above upper bound.
Consider an ergodic flow 
\(
\Phi_{\mathcal M}(A) = \int_{x\in A}{\mathcal M}(x,A^c)\,\pi(dx),
\) as the amount of total flow from $A$ to $A^c=\Theta\setminus A$. The conductance of $\mathcal M$ is a measurement of the bottleneck flow adjusted by the volume:
\(
\psi^*_{\mathcal M} := \inf_{A\subset \Theta,\pi(A)<1/2} \frac{\Phi_{\mathcal M}(A)}{\pi(A)}.
\)
The corrollary 3.3 of \cite{lovasz1992randomized} states that
$\sup_{A\in \Theta}| \mathcal M^t(x^0, A) -  \pi(A) | \le \sqrt{M} \{1-(\psi^*_{\mathcal M})^2/2 \}^t$, with $M = \sup_{A\subset \Theta} \pi^{0}(A)/\pi(A)$.

Therefore, when comparing two Markov chains, a large conductance $\psi^*_{\mathcal M}> \psi^*_{\mathcal M'}$ means that $\mathcal M$ has a faster-diminishing upper-bound rate on the total variation distance, hence a smaller upper-bound estimate on the mixing time, when compared with $\mathcal M'$. Although this is not a direct comparison between two mixing times, it offers theoretical insights into why one algorithm empirically shows a faster mixing of Markov chains than the other.

We now focus on the Markov chain generated by the baseline $\mathcal K(\theta^t, \cdot)$.
 For a sufficiently small $\epsilon>0$, we define an $\epsilon$-expansion from the infimum
\(
\mathcal A_\epsilon^*(\mathcal K) :=\biggl\{A\subset \Theta\ \bigg|\ \frac{\Phi_{\mathcal K}(A)}{\pi(A)} < \psi^*_{\mathcal K} + \epsilon ,\pi(A)<\frac{1}{2}  \biggr\}.
\)
We consider the graph-accelerated Markov chain with $\mathcal R= w \mathcal Q + (1-w)\mathcal K$ where the transition kernel $\mathcal Q$ has the same invariant distribution $\pi$. We have the following guarantee.

\begin{theorem}\label{lemma:general}
  If there exists a sufficiently small $\epsilon > 0$, for any $A \in \mathcal A_\epsilon^*(\mathcal K)$,  $\Phi_{\mathcal Q}(A)>\Phi_{\mathcal K}(A)$, then there exists $w\in (0,1]$ such that $\psi^*_{\mathcal R} > \psi^*_{\mathcal K} $.
\end{theorem}

The above result shows that $\mathcal Q$ only needs to improve the ergodic flow on $\mathcal{A}_\epsilon^*(\mathcal K)$ where the $\epsilon$-expansion of bottleneck flow of $\mathcal K$ happens. This means that as long as $\mathcal Q$ improves the flow in $\mathcal{A}_\epsilon^*(\mathcal K)$, the mixture kernel $\mathcal R$ will have potential acceleration.

\begin{remark}
Theorem \ref{lemma:general} is qualitative because we are limited to comparing  two upper bounds of mixing time. Nevertheless, the result  formalizes our comment in Remark \ref{rm:nonergodicity_of_q} --- we do not need $\mathcal Q$ alone to form a fast-mixing Markov chain. 
	As an intuitive example, for sampling a $k$-modal distribution via the mixture kernel $\mathcal K$, including jumps over a barebone graph with only $k$ nodes (each located near a unique mode) as $\mathcal Q$ will help improve the mixing of  Markov chains.
	\end{remark}

\subsection{Scaling of Acceptance Rate in High Dimension \label{sec:scaling_limit}}
For the specific relaxation distribution introduced in Section 2.3,
we give a theoretical characterization of the acceptance rate in terms of its rate of change as $p$ grows. For now, we treat the approximate sample size $m$ as a sufficiently large number that satisfies the two assumptions below, and we will discuss the associated requirement on $m$ later.

 Our goal is to obtain a lower bound on the expected acceptance rate $\mathbb{E}_{\theta^t \sim \Pi(\theta \mid y)} \alpha(\theta^t,\theta^*)$. Since $\mathbb{E}_{\theta^t \sim \Pi(\theta \mid y)} \alpha(\theta^t,\theta^*) \ge \mathbb{E}_{\theta^t \sim \Pi(\theta \mid y)} 1(\theta^t \in \mathcal B)\alpha(\theta^t,\theta^*)$ for $\mathcal{B}\subset \Theta$. We now find a $\mathcal B$ that could yield a tractable bound.
 
 We first exclude those $\beta^j: \min_{i\in B(j,r) \setminus j}  | \log \Pi(\beta^i \mid y) - \log\Pi(\beta^j \mid y)| > \kappa$ with $\kappa$ the chosen constant in \eqref{eq:mst}. For the remaining $\beta^j$'s, we can form an $\delta$-covering, denoted by $\tilde B_1$. That is,  for any $x\in \tilde B_1$, $\min_j \|x-\beta^j\|_2\le \delta$. We choose $\delta = c_2 p^{-c_3}$ with $c_2>0$ and $-\infty<c_3<1/2$.
Next, we assume (A1) there exists set $\tilde  B_2$ and $p$-independent constants $c_{1} > 0$ and $\gamma \ge 0$ such that for any $(\theta, \theta')\in \tilde   B_2\times \tilde  B_2$,
\(
|\log \Pi(\theta\mid y) - \log \Pi(\theta'\mid y)| \le  c_{1} p^\gamma \|\theta-\theta'\|,
\)
where $\|\cdot\|$ denotes some norm. This is commonly referred to as a $(c_{1} p^\gamma)$-smoothness condition \citep
{Bubeck_2015} if one uses Euclidean norm, and 
recently considered by \cite{tang2024computational} in the study of Metropolis-Adjusted Langevin algorithms. The difference here is that we only impose this condition on a subset $\tilde  B_2\subset \Theta$, hence the condition is relatively easy to satisfy. Taking $\mathcal B= \tilde B_1 \cap \tilde  B_2$, (A2) we assume $\mathcal B$ has posterior probability $\int_{\mathcal B} \Pi(\theta\mid y) d\theta= \mu_{\mathcal B}$ bounded away from zero. 

\begin{theorem}\label{eq:mh_acc_rate_lb}
	Under (A1) and (A2), we further assume $(b_i-a_i)/(b_j-a_j)<c_4$ 	 for all $i,j: i \in B(j;r)$, $|B(j;r)| \le c_5$ for all $j$, and $l\le \delta$. For the acceptance rate in \eqref{eq:detailed_balance}, we have
	\(
	\mathbb{E}_{\theta^t \sim \Pi(\theta \mid y)} \alpha(\theta^t,\theta^*) > \frac{\mu_{\mathcal B}  }{c_4c_5} e^{-\kappa}\exp\{ - 2 c_1 p^{(\gamma-c_3)} \}.
	\)
\end{theorem}

\begin{remark}
Therefore, with suitable $\gamma$ and $c_3<1/2$, we have the acceptance rate vanishing at a rate slower than $O\{ \exp(-\tilde cp)\}$.
We can obtain $\gamma \le 1/2$ for many commonly seen posterior densities. For example, for $\log \Pi(\theta \mid y) = - \theta' A\theta + o(\|\theta\|_2^2)$ with positive definite $A$, we can find a $\tilde B_2$ inside the ball $\{\theta:\|\theta\|_1\le a_1\sqrt{p}\}$ which implies $\|\theta\|_2\le \|\theta\|_1\le a_1\sqrt{p}$. In that case, we have $\gamma=1/2$ for the Euclidean norm.
\end{remark}

We now discuss the required size  $m$ on the approximate samples. Obviously, the larger $m$, the larger area $\tilde B_1$ and $\mu_\mathcal B$ will be. To more precisely characterize its dependency on $p$, and suggest choice for $c_3$, we can think of a high posterior probability polytope $\mathcal P=(\theta: \theta= k_0 \Sigma^{1/2}_0 x + a_0,  \|x\|_1\le 1 )$ with for some $a_0\in\Theta$, $\Sigma_0$ positive definite, and some fixed and dimension-independent $k_0>0$  so that $\mu_{\mathcal P} = \int_{\mathcal P} \Pi(\theta\mid y) d\theta \gg 0$.  Assuming the approximate sampler can generate points in $\mathcal P$, the key question is how many balls $(x: \|x-\theta_j\|\le \delta)$ are needed for covering $\mathcal P$?

The answer depends on the type of norm used in $\|x-\theta_j\|$. In the following, we consider using $\|x-\theta_j\|_{\Sigma_0} =  \sqrt{(x-\theta_j)^{T} \Sigma_0^{-1} (x-\theta_j)}$, which simplifies the problem to the covering a unit $L1$-ball using small $L2$-balls. The celebrated Maurey's empirical method \citep{pisier1999volume} shows that, to cover a unit $L1$-ball, we only need at least  $m=(2p+1)^{O(1/\delta_0^2)}$-many $\delta_0$-$L2$-balls with radius $\delta_0$, provided that $\delta_0 > p^{-1/2}$. With appropriate scaling, we reach the choice of $\delta = c_2 p^{-c_3}$, with $c_3<1/2$. {Substituting} into the lower bound of $m$, we see that $m= (2p+1)^{ O(p^{2c_3})}$. Therefore, we see that $c_3 = 0$ gives our suggested choice of $m=O(p)$, which balances between controlling acceptance rate {decay} and preventing excessive demand on the {number} of approximate samples.

\begin{remark}
In the above, we focus on a general high-dimensional setting with a high posterior probability set $\mathcal P$. On the other hand, in special but often encountered cases where the high posterior probability set can be found as a $\delta$-neighborhood of a $\tilde p$-dimensional polytope (with $\tilde p\ll p$, such as in sparse regression where most elements of $\theta$ are close to $0$), we can change the above paragraph to be based on a $\tilde p$-dimensional $L1$-ball. A further reduction of $m$ could be possible under additional assumptions on the $\tilde p$-dimensional polytope.
\end{remark}

The reason that we choose to study covering $\mathcal P$, an affinely transformed $L1$-ball, instead of an affinely transformed  $L2$-ball (ellipsoid), is that the covering number of the unit $L2$-ball with $\delta_0$-radius $L2$-balls (each of radius $\delta_0<1$) is $m=O(1/\delta_0)^p$ \citep{vershynin2015estimation}, which would be excessively large. On the other hand, since traditionally it is more common to think of a high posterior probability as an ellipsoid $\mathcal E=(\theta: \theta= k_0 \Sigma^{1/2}_0 x + a_0,  \|x\|_2\le 1 )$ than polytope $\mathcal P=(\theta: \theta= k_0 \Sigma^{1/2}_0 x + a_0,  \|x\|_1\le 1 )$, we want to give some characterization on the probability of $\mathcal P$. We focus on the case when  $(\theta \mid y)$ is a $p$-dimensional sub-Gaussian random vector \citep{vershynin2018high}. We say $x\in \mathbb{R}^p$ is a sub-Gaussian random vector, if $\mathbb{E}x=0$ and for any $v\in\mathbb{R}^p: \|v\|_2=1$, $v^{\rm T} x$ is sub-Gaussian with variance proxy $\sigma^2_0$, $\text{pr}(v^{\rm T}x \ge  d)\le 2\exp\{- d^2/(2\sigma^2_0)\}$ for any $d>0$. With transform $\theta = \Sigma^{1/2}_0 x + a_0$, $\theta$ is sub-Gaussian as well except with a different center and different variance proxy.
  It is not hard to see that if $x_{\theta}= \Sigma_0^{-1/2}(\theta-a_0)$ is sub-Gaussian random vector with variance proxy $\sigma^2_0$, then $\text{pr}( \|x_{\theta}\|_1 \le d) \ge 1- 2 \exp\{ -{d^2}/(2p\sigma^2_0)\}$. This can be obtained by observing $\|x\|_1= \tilde v'x$ for some $\tilde  v\in (-1,1)^p$ and $\|\tilde  v\|_2=\sqrt{p}$. Therefore, the associated $\mathcal P$ gives a high probability region.

\vspace*{-1cm}
\section{Simulations}
\vspace*{-0.5cm}
\subsection{Sampling Posterior with Non-convex Density Contour}
For sampling low-dimensional $\Pi(\theta\mid y)$, the random-walk Metropolis algorithm is appealing due to its low computational cost. 
 For low dimensional problems, a common choice for random walk proposal is $\No(\cdot ; \theta^t , s\, I)$, with $s>0$ the step size. {A} potential issue is that when the high posterior density region is not close to a convex shape, the step size $s$ would have to be small, {leading} to computing inefficiency. 
 The following example is often used as a challenging case \citep{haario1999adaptive}, with likelihood and prior
\(
y_i \stackrel{iid}\sim  \No (\theta^2_1+ \theta_2, 1^2), \text{ for } i=1,\ldots, n,
\qquad \theta \sim \No(0, I_2).
\)
{If the true parameters are chosen subject to the constraint} $\theta_1^2+\theta_2=1$, the posterior distribution of $(\theta_1,\theta_2)$ would spread around {the banana-shaped curve} $\{(\theta_1, \theta_2):\theta^2_1+ \theta_2 =1\}$.

\begin{figure}[H] \centering
                        \begin{subfigure}[t]{0.23\textwidth}
       \includegraphics[width=1\linewidth]{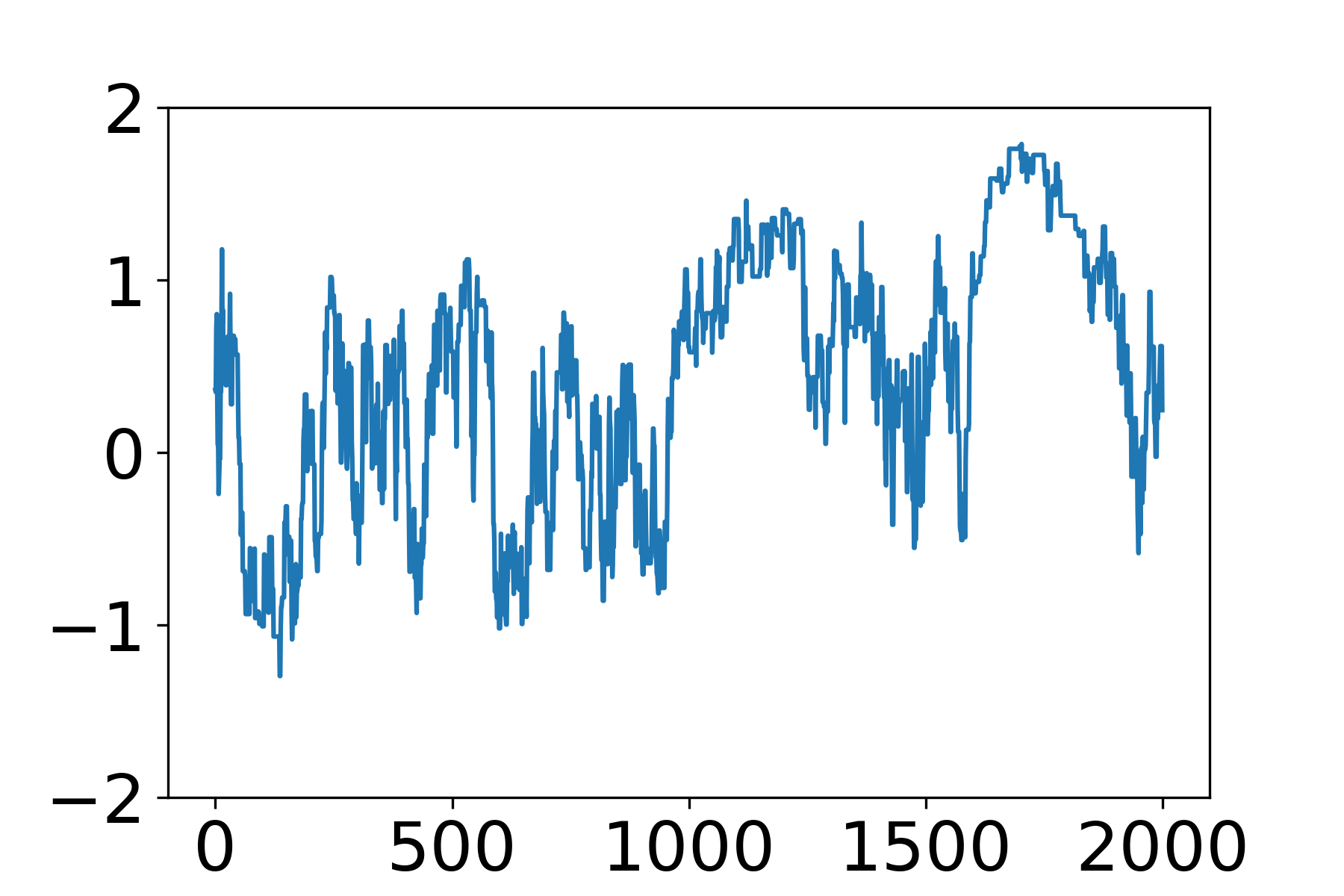}
\caption{\footnotesize  Traceplot of $\theta_1$ using random-walk Metropolis.}
 \end{subfigure}\;
 \begin{subfigure}[t]{0.23\textwidth}
 \centering
        \includegraphics[width=1\linewidth]{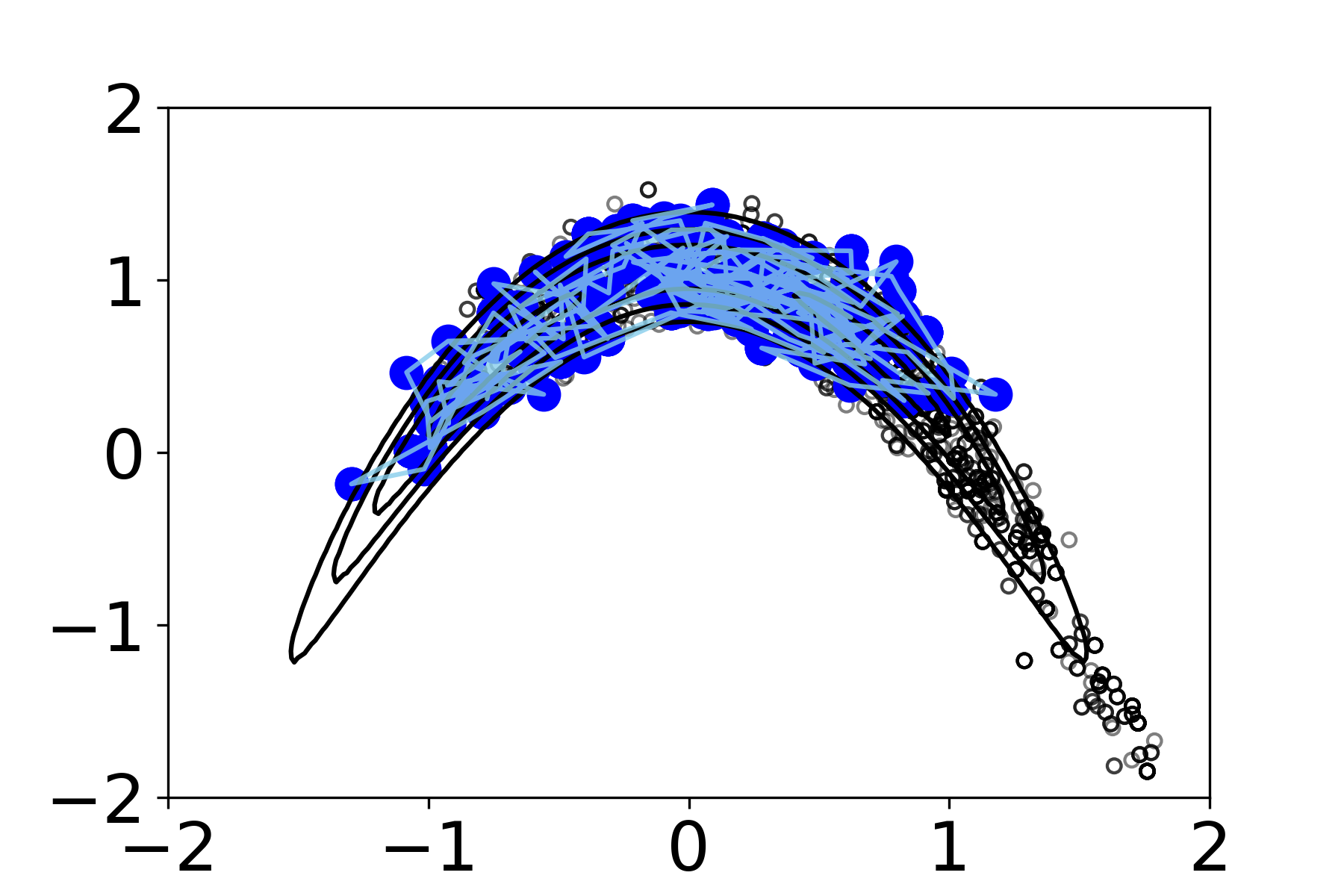}
\caption{\footnotesize  Markov chain sample of $(\theta_1,\theta_2)$ from the random-walk Metropolis. The first 400 sample points and traces are shown in blue.}
 \end{subfigure}\;
                         \begin{subfigure}[t]{0.23\textwidth}
 \centering
       \includegraphics[width=1\linewidth]{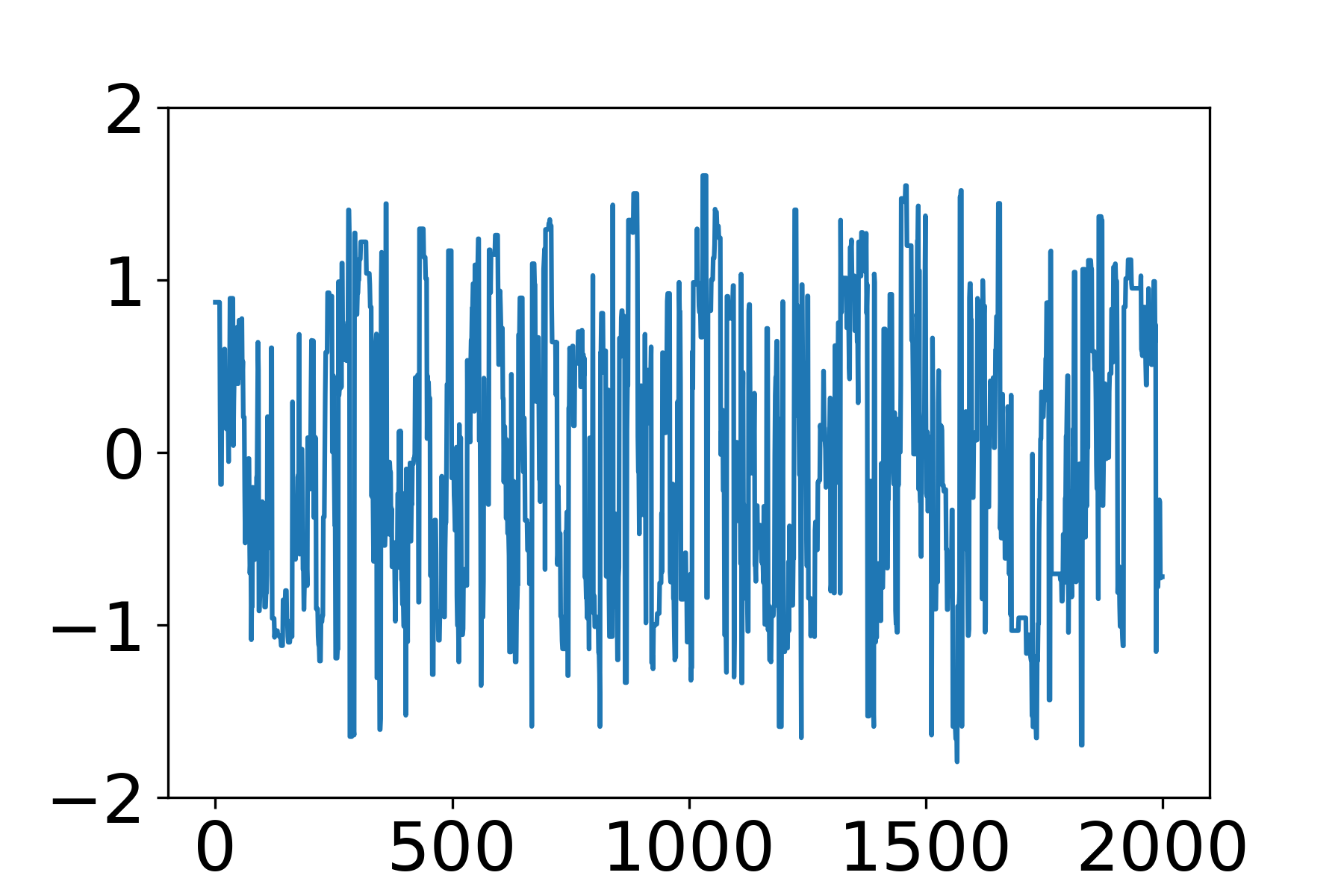}
\caption{\footnotesize  Traceplot of $\theta_1$ using the accelerated algorithm. }
 \end{subfigure}\;
   \begin{subfigure}[t]{0.23\textwidth}
 \centering
       \includegraphics[width=1\linewidth]{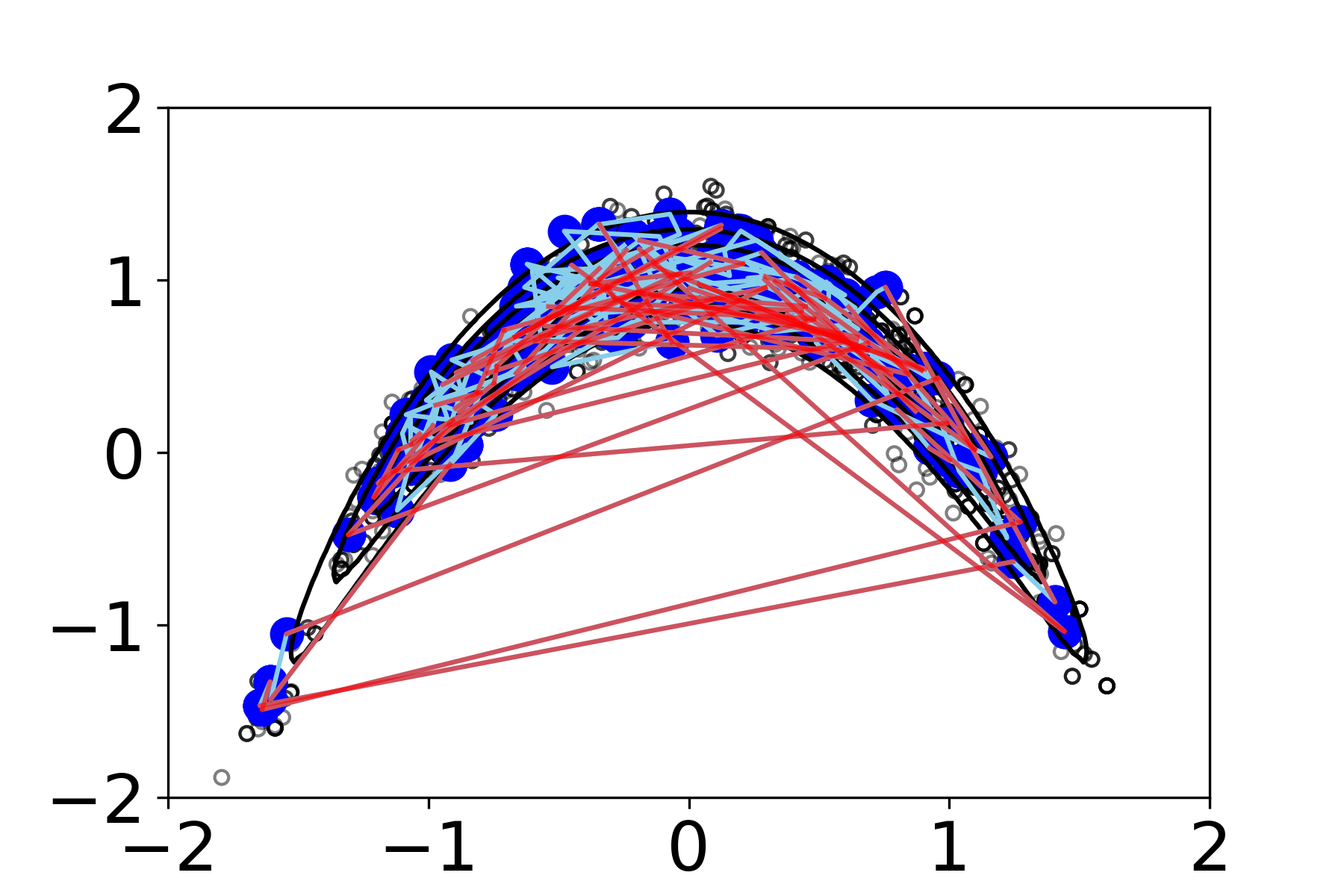}
\caption{\footnotesize  Markov chain sample of $(\theta_1,\theta_2)$ from the accelerated algorithm.  The first 400 sample points and traces are shown in blue, with successful graph jump steps shown in red.}
                       \end{subfigure}
  \caption{Graph-accelerated random-walk Metropolis for sampling a posterior distribution of banana shape. \label{fig:banana}}
 \end{figure}

Using random-walk Metropolis as the baseline algorithm, we tweak $s$ to around $0.5$ so that the Metropolis acceptance rate is around 0.234. We run the algorithm for 3000 iterations and use the last 2000 as a Markov chain sample. Figure \ref{fig:banana}(a)(b) shows that it takes a long time for the sampler to move from one end to the other.

For acceleration, we first obtain $100$ approximate samples from a variational method based on a 10-component Gaussian mixture $\sum_{k=1}^{10} \tilde w_k \No(\tilde \mu_k, I \tilde\sigma^2)$, then we run the accelerated algorithm. As shown in Figure \ref{fig:banana},  the accelerated algorithm jumps rapidly between the two ends, leading to improved mixing performance. The effective sample size per iteration for  $\theta_1$ from the baseline algorithm is 0.16\%, while the one for the accelerated version is 6.1\%. 

\vspace*{-1cm}
\subsection{Numerical Results on the Change of Acceptance Rate}
In Section 3, we gave a lower-bound quantification of the Metropolis-Hastings acceptance rate under a theoretical setting with increasing dimensions. To show empirical evidence that the acceptance rate remains positive and away from zero in practical settings, we conduct simulations under different dimensions $p$ and approximate sample sizes $m$. Due to the page constraints, we provide the details in the Supplementary Materials S.3.

\vspace*{-1cm}
\section{Application: Estimating Latent Gaussian Model for Power Outage Data}

To show that our algorithm works well in relatively large dimensions, we experiment with a latent Gaussian model for count data. Specifically, we use the power outage count for a zip code area in the south of Florida collected during a 90-day time period in the 2009 hurricane season. There are $n=513$ records of outage counts $y_i \in \mathbb{Z}_{\ge 0}$, reported at irregularly-spaced time points. To ease the prior specification, we first rescale the time records to be in $[0,1]$, and denote the transformed time by $t_i$. We use the following likelihood based on a negative-binomial latent Gaussian model, with latent Gaussian covariance $\Sigma_{i,j}(\tau,h)= \tau \exp[ - (t_i-t_j)^2/{2h}]$, leading to augmented likelihood:
\(
L(y,z \mid \tau,h ) =  (2\pi)^{-n/2}|\Sigma(\tau,h)|^{-1/2}\exp \bigg[ - \frac{1}{2} z^{\rm T} \{\Sigma(\tau,h)\}^{-1} z \bigg]\prod_{i=1}^{n}\frac{\exp(r z_i)}{ \{ 1+\exp(z_i)\}^{r+y_i}} .
\)
For prior specification, we use $h\sim \text{Inverse-Gamma}(2,1)$ for the bandwidth $h>0$, $r\sim \text{No}_{0,\infty)}(0,1)$ for the inverse dispersion parameter $r>0$, $\tau\sim \text{Inverse-Gamma}(2,1)$ for the scale $\tau>0$.

\begin{figure}[H] \centering
                        \begin{subfigure}[t]{0.24\textwidth}
       \includegraphics[width=1.05\linewidth]{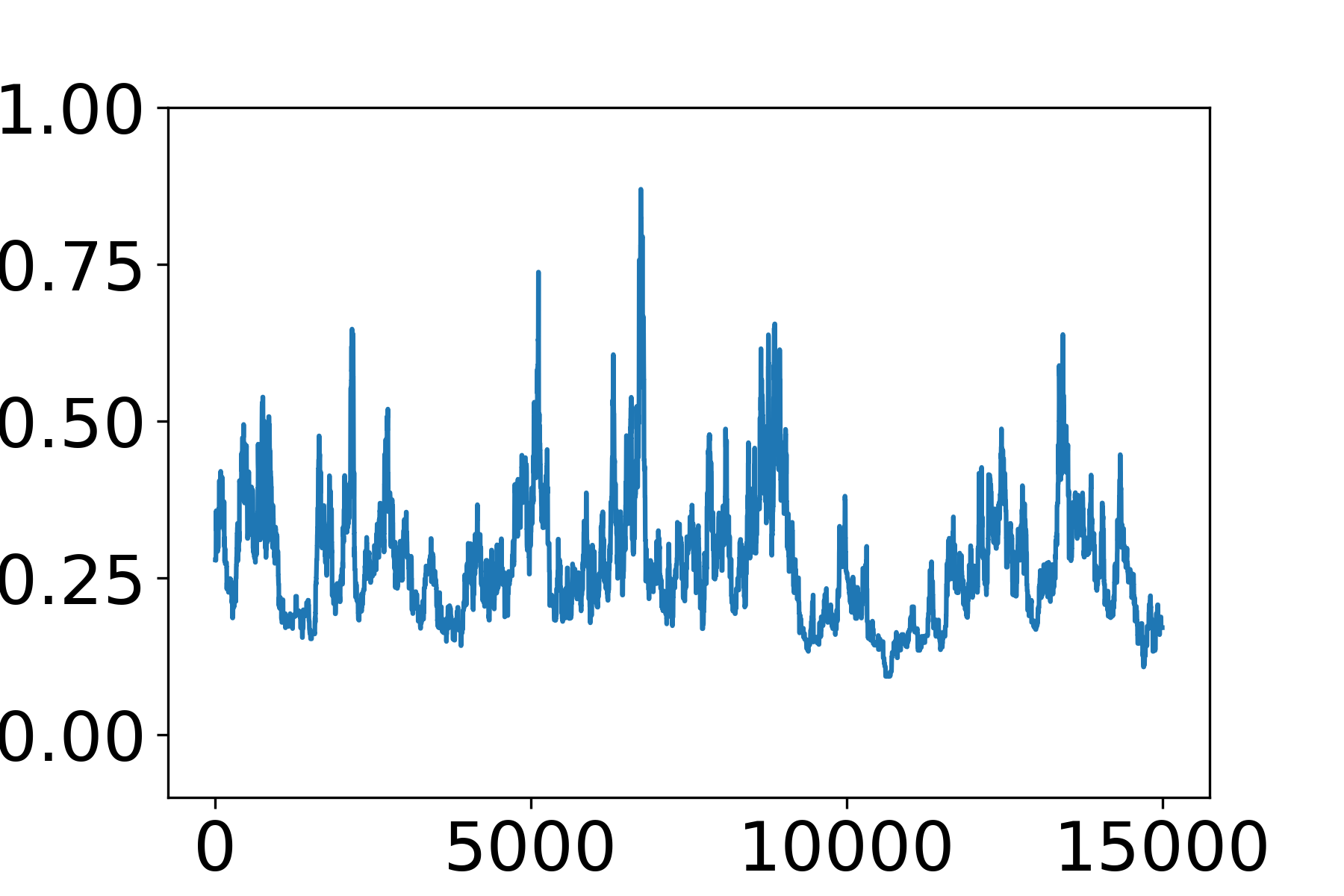}
\caption{\footnotesize  Traceplot of $h$ produced by the  Gibbs sampler.}
 \end{subfigure}
                         \begin{subfigure}[t]{0.24\textwidth}
       \includegraphics[width=1.05\linewidth]{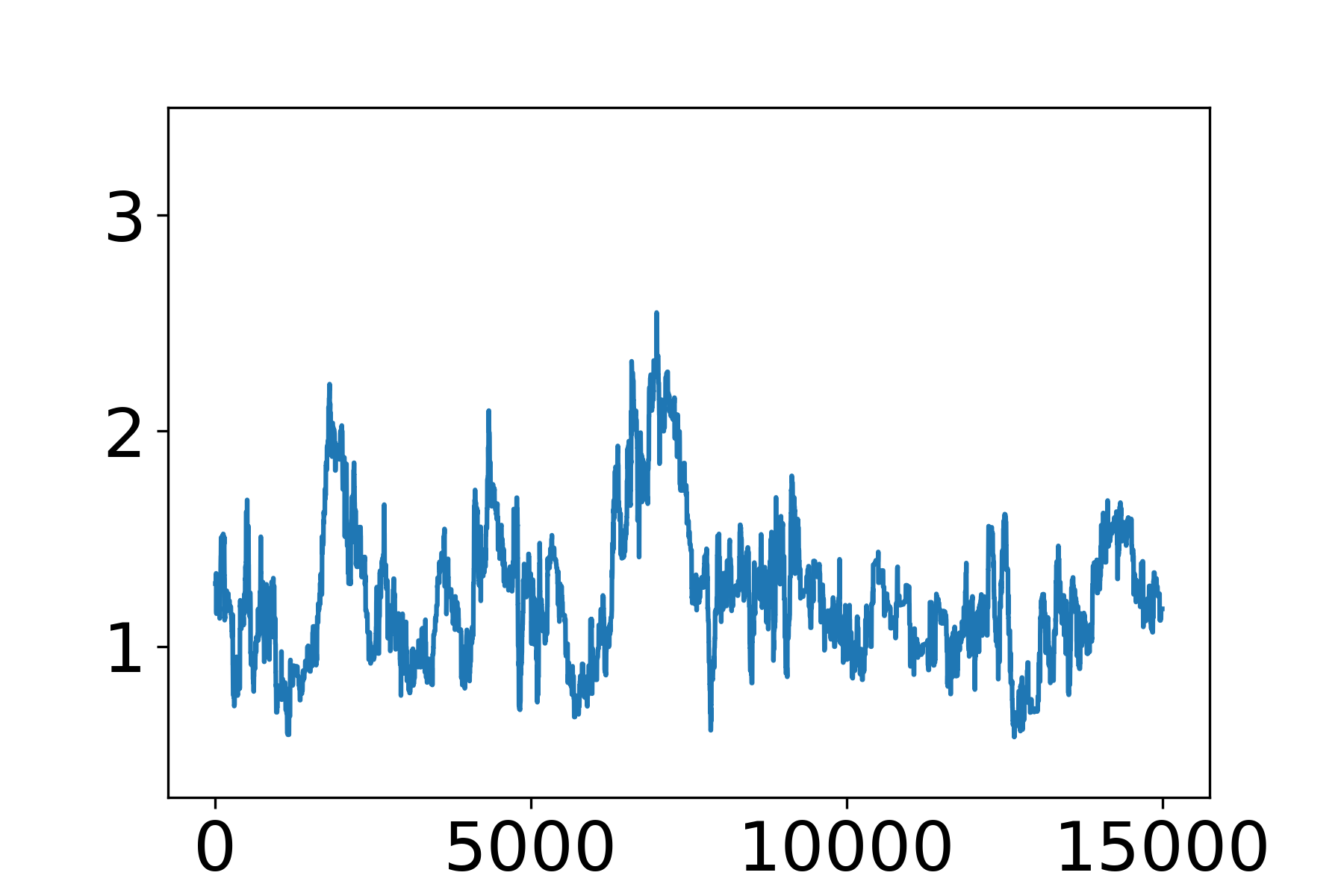}
\caption{\footnotesize  Traceplot of $r$  produced by the  Gibbs sampler.}
 \end{subfigure}
                         \begin{subfigure}[t]{0.24\textwidth}
 \centering
       \includegraphics[width=1.05\linewidth]{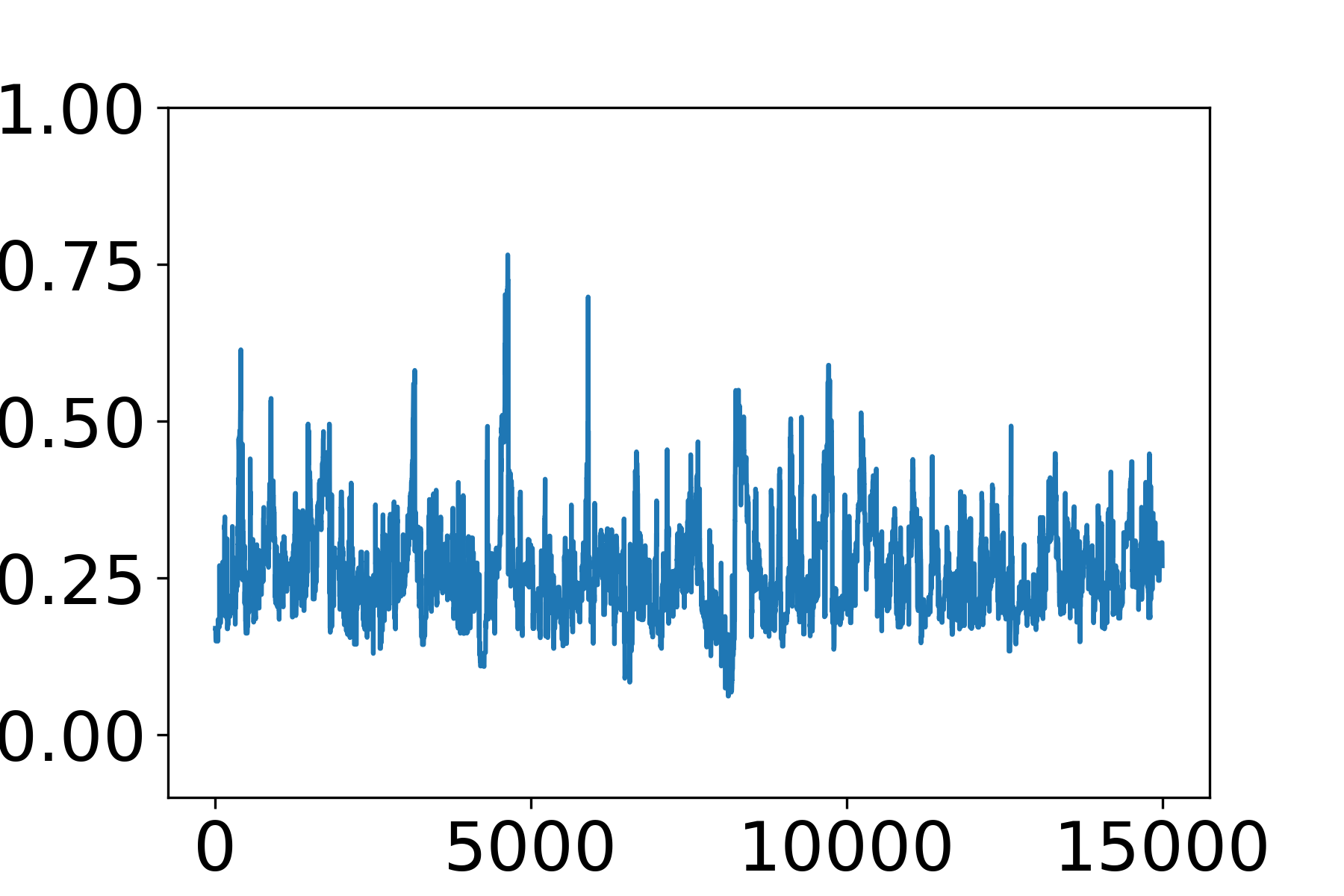}
\caption{\footnotesize  Traceplot of $h$  produced by the accelerated sampler. }
 \end{subfigure}
   \begin{subfigure}[t]{0.24\textwidth}
 \centering
       \includegraphics[width=1.05\linewidth]{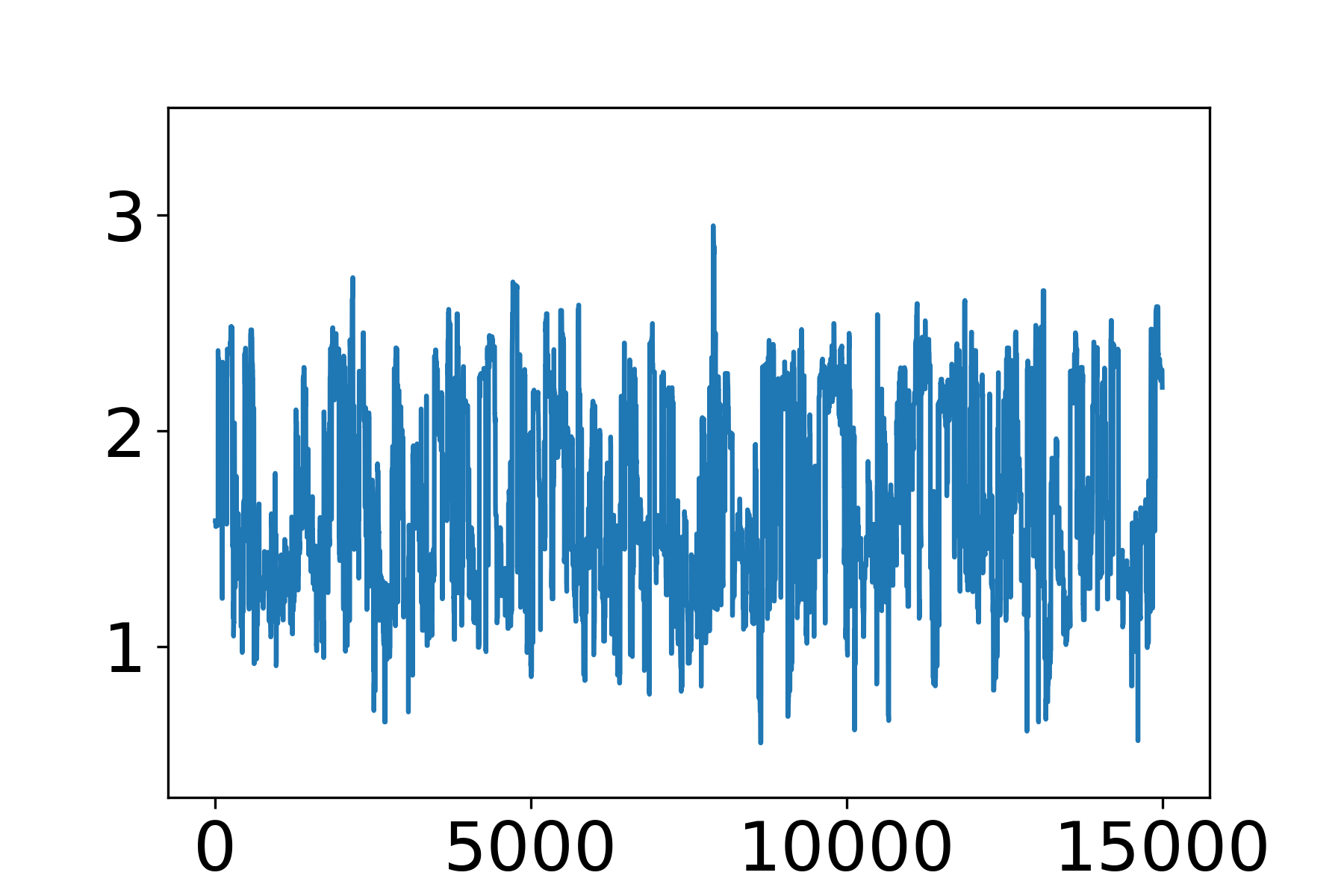}
\caption{\footnotesize  Traceplot of $r$  produced by the accelerated sampler. }
 \end{subfigure}
     \begin{subfigure}[t]{0.47\textwidth}
     \begin{overpic}[width=1.1\linewidth]{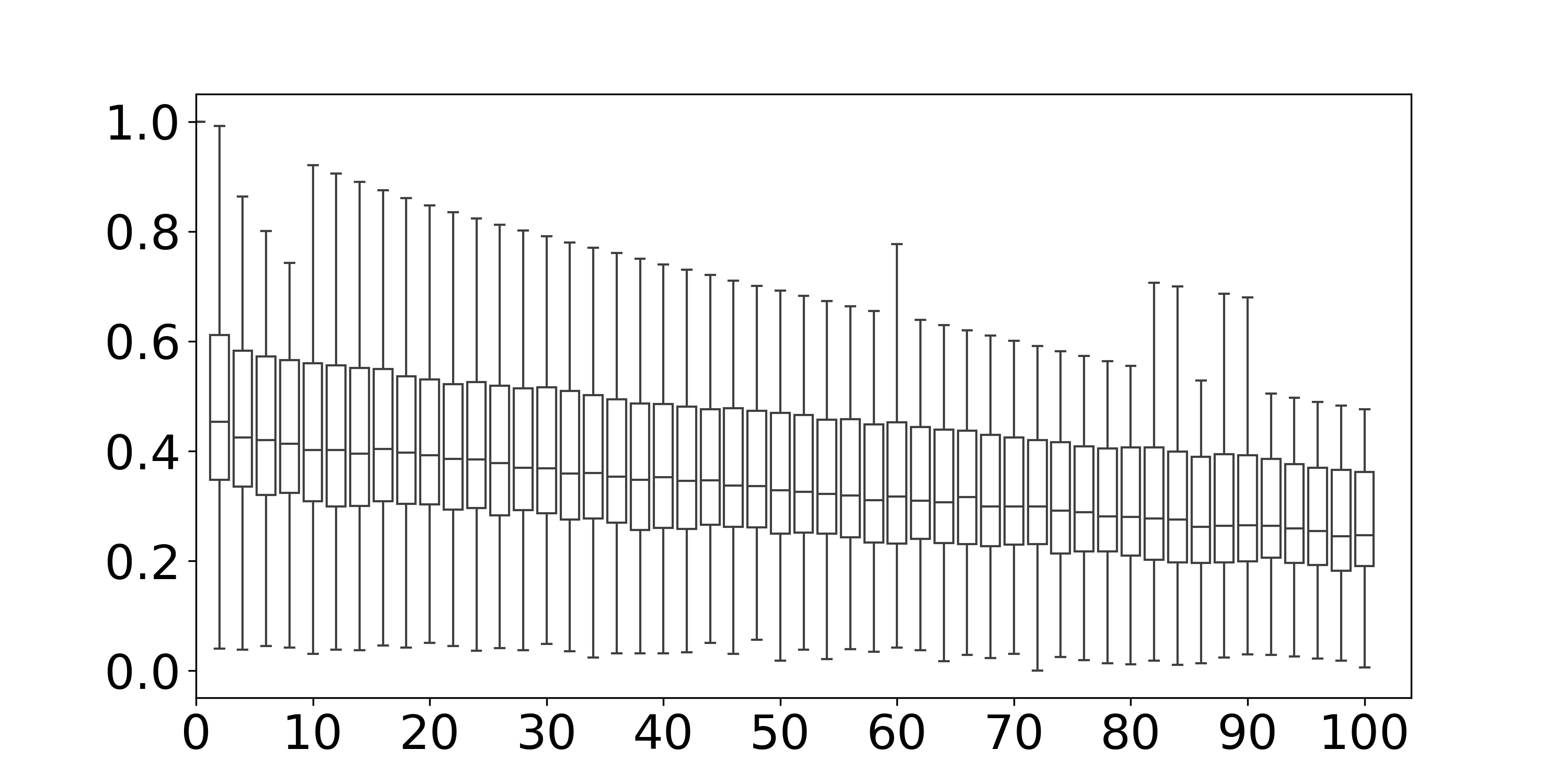}
    \put(50, -2){\scriptsize Lag}
    \put(-1,12){\rotatebox{90}{\scriptsize Autocorrelation}}
    \end{overpic}
\caption{\footnotesize  Autocorrelation plot for all parameters and latent variables, from the samples produced by Gibbs sampler.}
 \end{subfigure}\;
    \begin{subfigure}[t]{0.47\textwidth}
 \centering
     \begin{overpic}[width=1.1\linewidth]{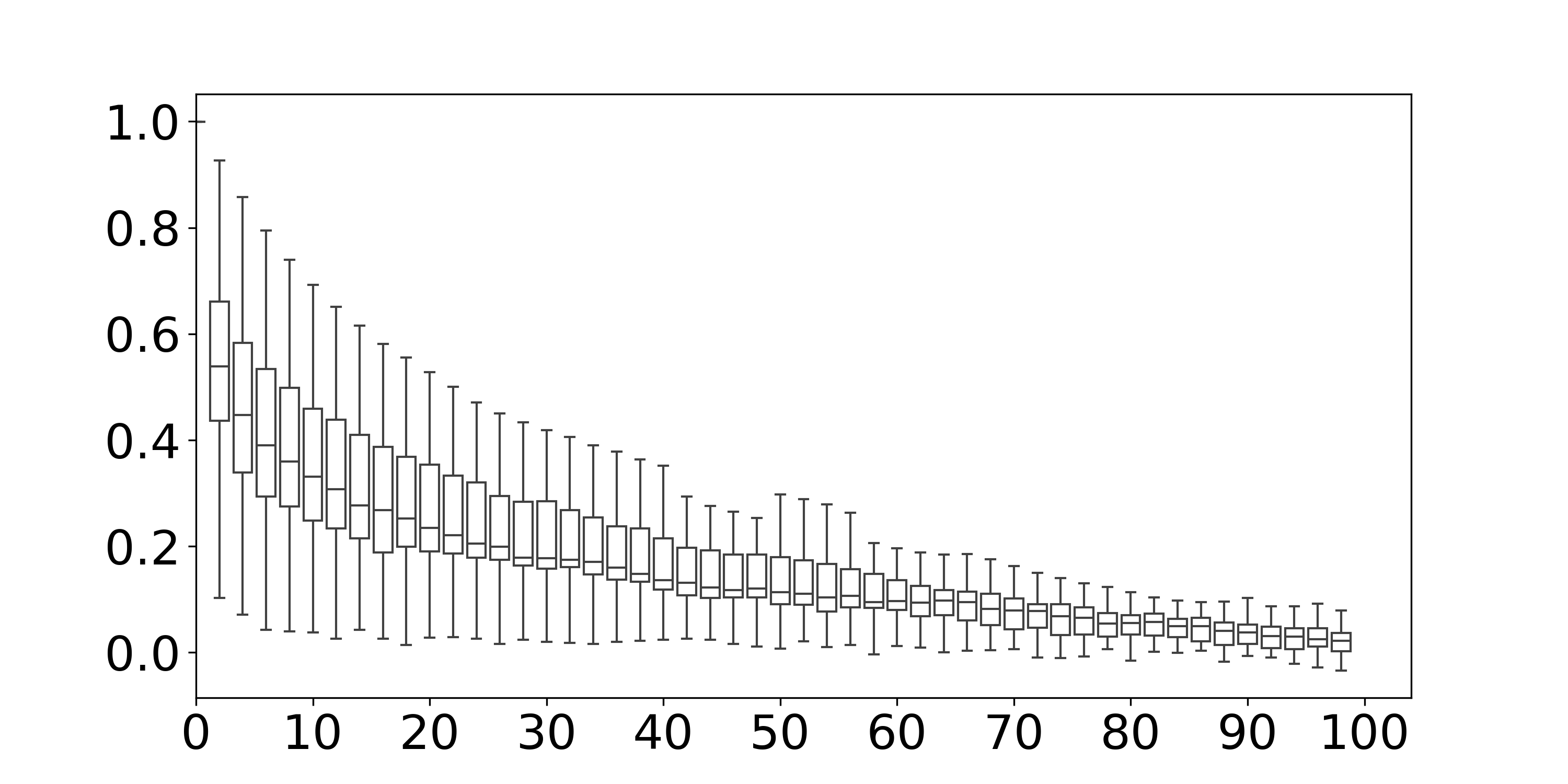}
    \put(50, -3){\scriptsize Lag}
    \put(-1,12){\rotatebox{90}{\scriptsize Autocorrelation}}
    \end{overpic}\caption{\footnotesize  Autocorrelation plot for all parameters and latent variables, from the samples produced by accelerated sampler.}
 \end{subfigure} 
    \begin{subfigure}[t]{0.28\textwidth}
 \centering
            \begin{overpic}[height=1.3in,width=0.8\linewidth]{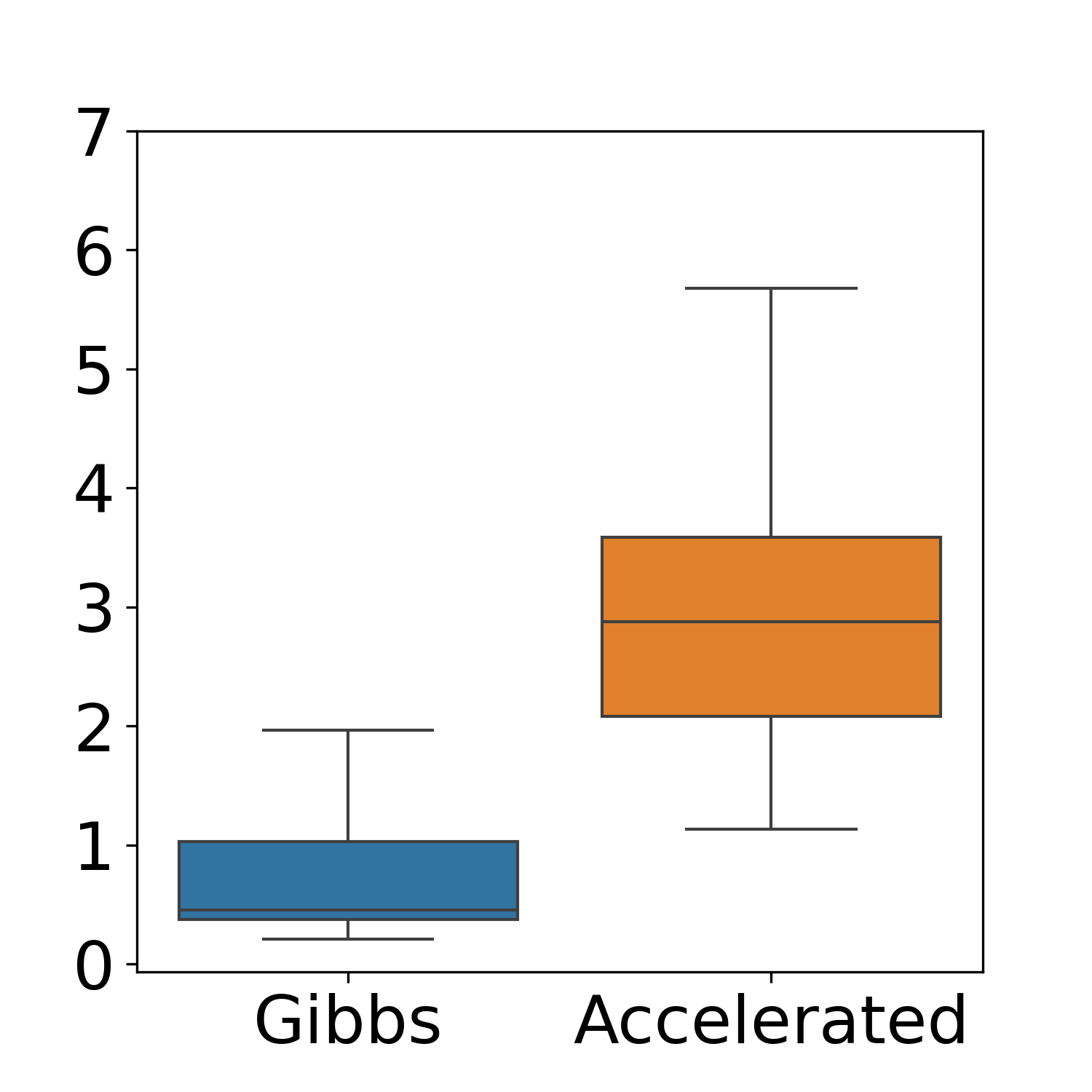}
    \put(-3,12){\rotatebox{90}{\scriptsize Effective sample size \%}}
    \end{overpic}
\caption{\footnotesize Boxplots showing effect sample sizes per iteration for two algorithms.}
 \end{subfigure}\;
   \begin{subfigure}[t]{0.32\textwidth}
 \centering
     \begin{overpic}[width=1.05\linewidth]{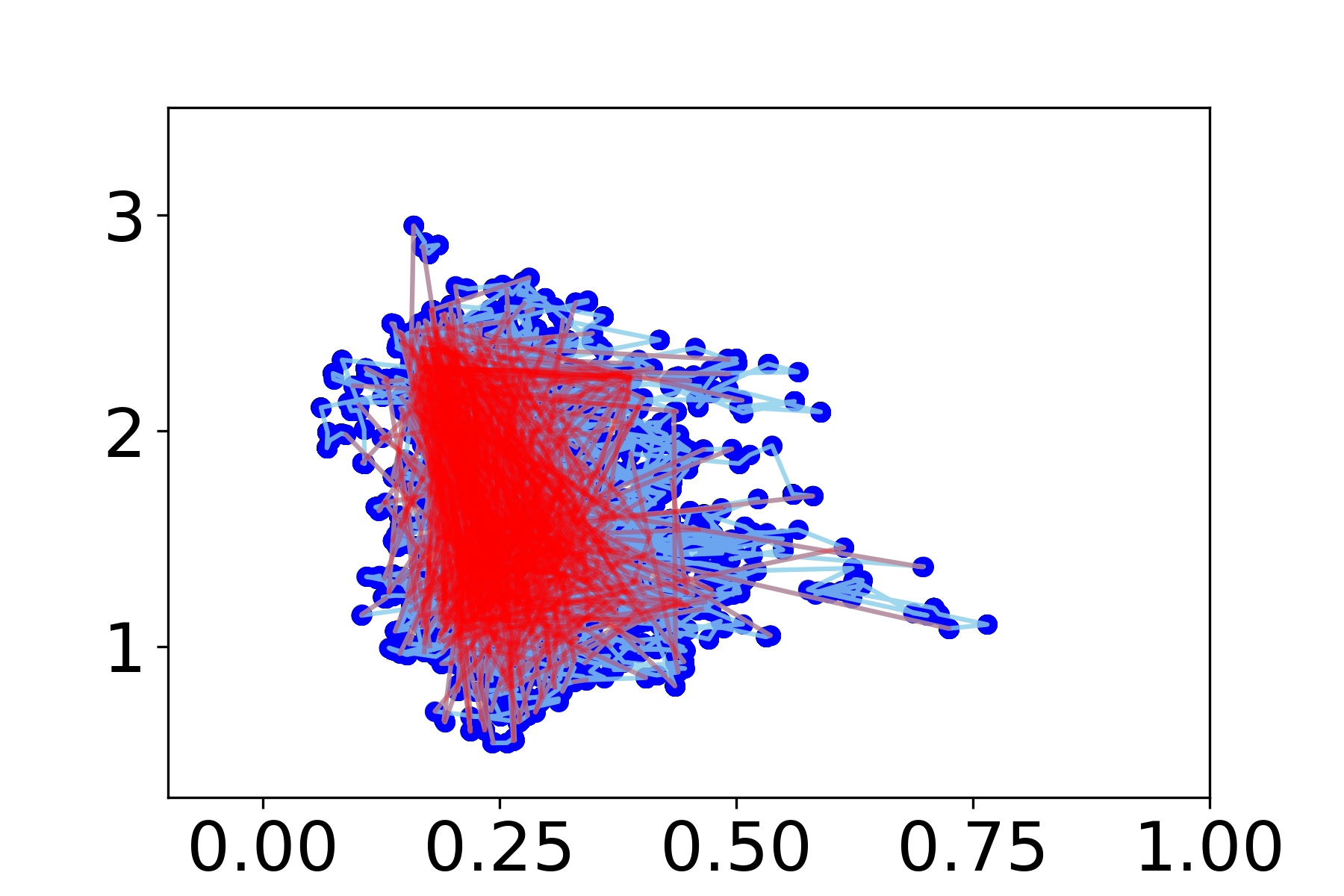}
    \put(50, -3){\scriptsize $h$}
    \put(,30){\rotatebox{90}{\scriptsize $r$}}
    \end{overpic}
\caption{\footnotesize  Markov chain sample of $(h,\tau)$ from the accelerated algorithm, with successful graph jumps shown in red.}
                       \end{subfigure}\;
     \begin{subfigure}[t]{0.32\textwidth}
      \begin{overpic}[width=1.05\linewidth]{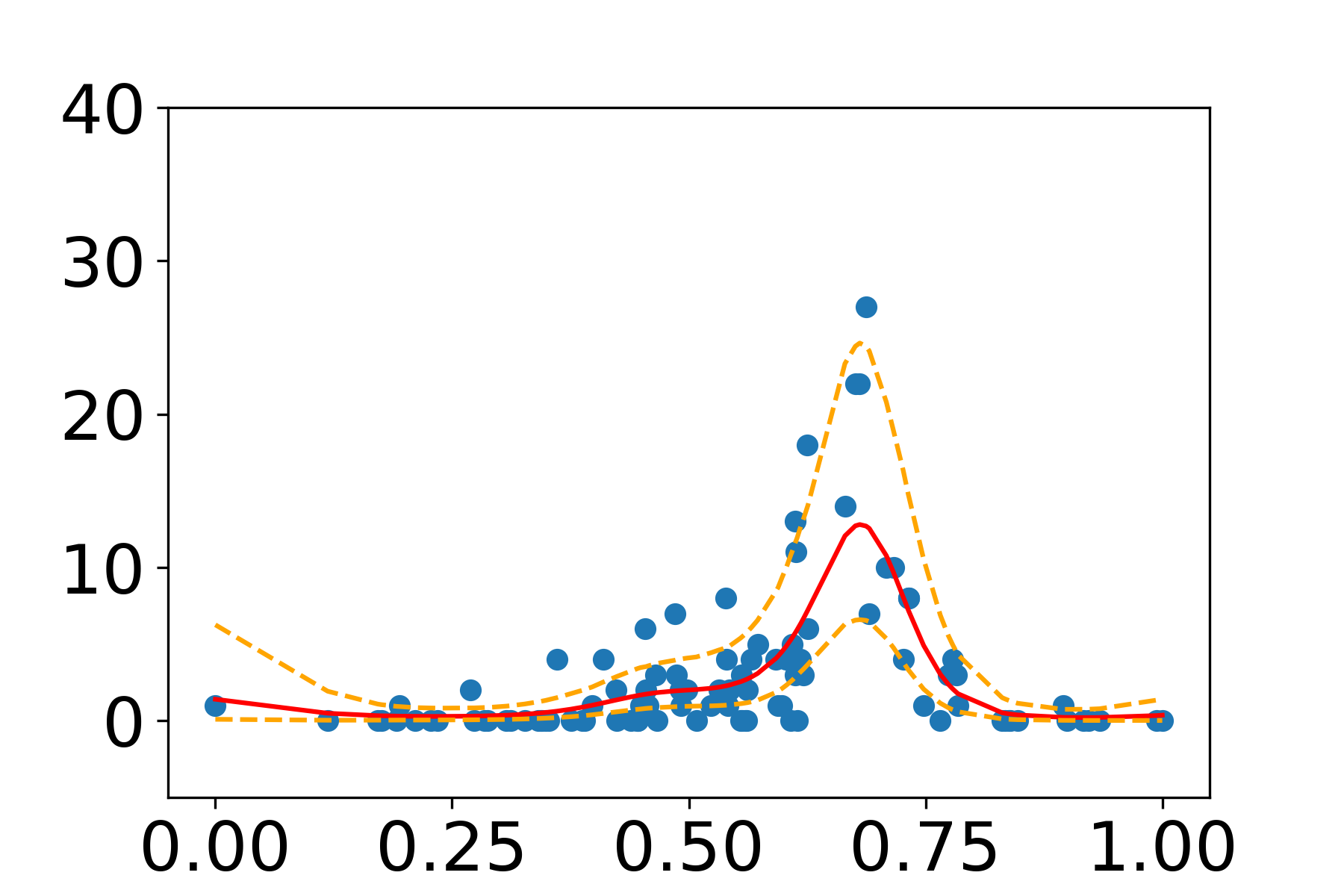}
    \put(50, -3){\scriptsize $t$}
    \put(-2,30){\rotatebox{90}{\scriptsize $y$}}
    \end{overpic}
\caption{\footnotesize  Posterior estimate of the mean curve using the accelerated algorithm.}
                       \end{subfigure}
  \caption{Sampling a posterior distribution of a latent Gaussian model for count data. \label{fig:lgm}}
 \end{figure}

We first describe the baseline algorithm posterior sampling. 
Using P\'olya-Gamma latent variable $\omega_i$ \citep{polson2013bayesian}, denoted by $\omega_i\sim \text{PG}(\cdot \mid r+y_i ,0)$, we have
\(
\frac{\{\exp(z_i)\}^r}{ \{ 1+\exp(z_i)\}^{r+y_i}} = 2^{-(r+y_i)}
\exp\{(\frac{r-y_i}{2}) z_i\}
\int 
\exp(- \omega_i  z^2_i/2) \text{PG}(\omega_i \mid r+y_i ,0)  \textup d \omega_i.
\)
 We have closed-form updates for most of the latent variables and parameters, 
$\omega_i \sim \text{PG}(r+y_i, z_i)$ for $i=1,\ldots, n$, 
$z\sim \text{No} [
\{ \Sigma^{-1}+ \text{diag}(\omega_i) \}^{-1}\{(r-y)/2\}, 
\{ \Sigma^{-1}+ \text{diag}(\omega_i) \}^{-1} 
]$ and $\tau \sim \text{Inverse-Gamma} \{n/2+2,  z^{\rm T} \tilde\Sigma^{-1}(h)z/2 +1\}$ with $\tilde\Sigma_{i,j}(h) = \exp[ - (t_i-t_j)^2/{2h}]$. On the other hand, since $h$ and $r$ do not have full conditional distribution available in closed form, we use softplus reparameterization $h=\log\{ 1+\exp(\tilde h)\}$ and $r=\log\{1+\exp(\tilde r)\}$ and use random-walk Metropolis algorithm with proposal $\No \{ \cdot ; (\tilde h,\tilde r)^t , I s\}$ to obtain an update on $(\tilde h, \tilde r)\in \mathbb{R}^2$, then transform to $(h,r)$.  In the random-walk Metropolis algorithm, we use the posterior with $(\tau,\omega)$ integrated out, and tweak $s$ so that the acceptance rate is around 0.234. We run the baseline algorithm for 20,000 iterations, and treat the first 5,000 as burn-in. As shown in Figure \ref{fig:lgm}(a)(b) and (e), the baseline Gibbs sampling algorithm suffers from critically slow mixing. Even at the 100-th lag, most of the parameters and latent variables still show autocorrelations near 40\%.

For the acceleration algorithm, we obtain approximate samples of $\beta^j=(z,\tau,h,r)^j\in \mathbb{R}^{103}$ by simply taking the first $1,000$ Markov chain samples after the burn-in period from the slow-mixing Gibbs sampler, then we construct the graph and run the accelerated algorithm for 20,000 iterations (with the first 5,000 as burn-in). Despite the relatively large dimension, the graph jump steps had $18.4\%$ of success rate \{red lines in Figure \ref{fig:lgm}(h)\}. As shown in Figure \ref{fig:lgm}(c)(d) and (f), the accelerated algorithm leads to much-improved mixing performance. Almost all parameters and latent variables have autocorrelations reduce to small values after the 60-th lag. We plot the posterior mean curve  $r\exp(-z_i)$, and the point-wise 95\% credible band in Figure \ref{fig:lgm}(i).

\section{Discussion}
In our accelerated algorithm, we treat the graph as a fixed object. An interesting extension to explore is to allow the graph to keep growing, by adding some samples collected from the Markov chain. This idea is especially appealing in the sense that a {\em chain graph} is in fact a special tree graph without branches, which suggests opportunities to develop new algorithms such as {\em Markov tree Monte Carlo}. On the other hand, a critical issue is that the growing of tree graph would break the detailed balance condition, hence risking a failure of convergence to the target posterior distribution. One possible solution is to employ the well-known diminishing adaptation strategy \citep{roberts2009examples}, by making the differences between proposal kernels vanish as the iteration increases. Another possibility is to forsake the pursuit of detailed balance condition, but to satisfy some weaker and sufficient conditions that ensure global balance. Several non-reversible algorithms have been developed \citep{sohl2014hamiltonian,bierkens2016non} under different context, although how to extend the ideas to a growing graph remains an open question.

The accelerated algorithm described in this article can be generalized to the posterior sampling of discrete or combinatorial parameters. Nevertheless, choosing a relaxation distribution in high-dimensional discrete space can be challenging.  This issue could be potentially circumvented using continuous embedding as considered by several diffusion-based algorithms \citep{pakman2013auxiliary,nishimura2020discontinuous}.


\appendix

\section{Proof of Theorems}

\subsection{Proof of Theorem 1}
\begin{proof} 
To verify the detailed balance, it suffices to check the case when $\theta^{t+1}= \theta^*$. For any $\theta^{t}$ and $\theta^*$,
\(
& \Pi(\theta^t\mid y) \mathcal  Q (\theta^{t},  \theta^*)= \sum_{i \in B \{  \mathbb{N}(\theta^t);r \}}
\Pi(\theta^t\mid y)  | B (  j;r )|^{-1}
   F(\theta^* \mid \beta^i ,\theta^t) \alpha(\theta^t,\theta^*)\\
 & = \sum_{i \in B (  j;r )} 1 \big [\mathbb{N}(\theta^*)=i  \big] \\
 & \qquad \times
\min\bigg [ 
 { \Pi(\theta^*\mid y)  | B  (i;r ) |^{-1}
        F (\theta^t \mid \beta^{ j},\theta^*)
        }, 
  { \Pi(\theta^t\mid y)  | B (  j;r )|^{-1}
   F(\theta^* \mid \beta^i ,\theta^t) } \bigg ] \\
   & \stackrel{(a)}= 1 \big [\mathbb{N}(\theta^*)=i  \big] 1 \big [\mathbb{N}(\theta^t)=j  \big]\\
 & \qquad \times
\min\bigg [ 
 { \Pi(\theta^*\mid y)  | B  (i;r ) |^{-1}
        F (\theta^t \mid \beta^{ j},\theta^*)
        }, 
  { \Pi(\theta^t\mid y)  | B (  j;r )|^{-1}
   F(\theta^* \mid \beta^i ,\theta^t) } \bigg ] 
\
\)
where $(a)$ uses the almost sure uniqueness of projection, so that there is only one $i: 1[\mathbb{N}(\theta^*)=i]\neq 0$ at given $\theta^*$, and the fact that $1[\mathbb{N}(\theta^t)=j]=1$. Clearly, the last line is symmetric in $(\theta^t,\theta^*)$, hence $\Pi(\theta^t\mid y) \mathcal  Q (\theta^{t}, \theta^{*}) = \Pi(\theta^{*}\mid y) \mathcal  Q (\theta^{*}, \theta^{t})$.
\end{proof}

\subsection{Proof of Theorem 2}

\begin{proof} We consider the conductance under two cases:

\textbf{1) Transitioning from $A\in \mathcal A_\epsilon^*(\mathcal K)$:}

For any $A\in\mathcal A_\epsilon^*(\mathcal K)$, we have for any $w\in(0,1]$:
\[
\begin{aligned}
\frac{\Phi_{\mathcal R}(A)}{\pi(A)} = \frac{w\Phi_{\mathcal Q}(A)+ (1-w) \Phi_{\mathcal K}(A)}{\pi(A)} > \frac{\Phi_{\mathcal K}(A)}{\pi(A)} \geq \psi^*_{\mathcal K}.
\end{aligned}
\]

\textbf{2) Transitioning from $B\in \Theta \setminus\mathcal A_\epsilon^*(\mathcal K)$:}

For any $B\in  \mathcal B = \{ A\in  \Theta: \pi(A)<1/2, A\not \in \mathcal A_\epsilon^*(\mathcal K)\}$, we have $\Phi_{\mathcal K}(B)/\pi(B) \geq \psi_{\mathcal K}^* + \epsilon$.

Let $m_{\mathcal B}:= \inf_{B \in  \mathcal B} \Phi_{\mathcal Q}(B)/ \Phi_{\mathcal K}(B)\ge 0$, and for any $w\in (0,1]$ such that:
\[
w(1- m_{\mathcal B}) < \frac{\epsilon}{\psi^*_{\mathcal K} + \epsilon},
\]
we have
\[
\begin{aligned}
&    \frac{\Phi_{\mathcal R}(B)}{\pi(B)} =
    \frac{w\Phi_{\mathcal Q}(B) + (1-w) \Phi_{\mathcal K}(B)}{\pi(B)}
    =
\{ w\Phi_{\mathcal Q}(B)/\Phi_{\mathcal K}(B) + (1-w) \} \frac{\Phi_{\mathcal K}(B) } {\pi(B)} \\
& \ge \{ w m_{\mathcal B} + (1-w) \} \frac{\Phi_{\mathcal K}(B) } {\pi(B)}
     >  \frac{\psi^*_{\mathcal K} \Phi_{\mathcal K}(B)}{(\psi_{\mathcal K}^* + \epsilon)\pi(B)}
     \geq \psi^*_{\mathcal K}.
\end{aligned}
\]
To show that such a $w$ always exists, as well as choosing a large value for $w$: when $m_{\mathcal B}\ge 1$, we can choose $w=1$; when $m_{\mathcal B}<1$, we can choose $w = (1-m_{\mathcal B})^{-1}{\epsilon}/(\psi^*_{\mathcal K} + \epsilon) -\eta$, with $\eta>0$ sufficiently small so that $w>0$.

Combining 1) and 2), we see that there exists $w\in (0,1]$, such that $\psi^*_{\mathcal R} > \psi^*_{\mathcal K}$.
\end{proof}

\subsection{Proof of Theorem 3}
\begin{proof} The acceptance rate under our specified $F(\theta^* \mid \beta^i,\theta^t)$ is
\(\alpha(\theta^t,\theta^*)=
\min \bigg \{
  1, \frac{ \Pi(\theta^*\mid y)  | B ( i;r )|^{-1}(b_j-a_j)^{-1}
        }
  { \Pi(\theta^t\mid y)  | B (  j;r )|^{-1} (b_i-a_i)^{-1}
   } \bigg \}.
\)
We see that $\min_j\|\theta^*-\beta^j\|\le \delta$ by the way we generate $\theta^*$, hence $\theta^*\in \mathcal B$. Consider  any $\theta^t \in \mathcal {B}$, 
  \(
& \log  \Pi(\theta^*\mid y) - \log \Pi(\theta^t\mid y)\\
& \ge  \log  \Pi(\theta^*\mid y) - \log  \Pi(\beta^i\mid y)
+ \log  \Pi(\beta^j\mid y) - \log  \Pi(\theta^t\mid y)
- |\log  \Pi(\beta^i\mid y)-\log  \Pi(\beta^j\mid y)|\\
& \ge  - c_{1} p^\gamma ( \|\theta^*-\beta^i\| + \|\theta^t-\beta^j\|) 
-  \kappa \\
& \ge - 2 c_{1} p^\gamma\delta -  \kappa.
\)
Since we know $B (j;r ) \ge 1$, we have $| B (j;r )|/| B ( i;r )| \le c_5$. Including the bound ratio $(b_i-a_i)/(b_j-a_j)<c_4$, and taking expectation over $\theta^t\sim \Pi(\theta\mid y)$ yields the result.
\end{proof}

\section{Optimization Algorithm for Further Improvement on Graph Choice}
We provide the details on the optimization of a random walk transition probability matrix $P$.
One solution is using the dual ascent algorithm.
The minimization of spectral norm is equivalent to:
\(
 \min_{P,s}  & \; s\\
 \text{subject to } & \|P- 1_m \pi^{\rm T}_\beta\|_2\le s, \; s \ge 0\\
 & P1_m=1_m,
 \qquad \pi_\beta^{\rm T} P = \pi^{\rm T}_\beta,\\
&  P_{i,j}=0 \text{ if } (i\to j)\not \in E_{\bar G}, \qquad P_{i,j}\ge 0
\)
Using semi-definite programming, we can set up the Lagrangian:
\(
\mathcal L(P, Z, s, u,v, Y,\lambda) =
&s -
\text{tr} \bigg\{
\begin{bmatrix}
 Z_{11} &  Z_{12}	 \\
Z^{\rm T}_{12} & Z_{22}
\end{bmatrix}
\begin{bmatrix}
 sI &  (P- 1_m \pi^{\rm T}_\beta) 	 \\
 (P- 1_m \pi^{\rm T}_\beta) ^{\rm T} & sI	
\end{bmatrix}\bigg\}\\
&
+  u^{\rm T} (P1_m-1_m) +  (\pi_\beta^{\rm T} P - \pi^{\rm T}_\beta) v
- \text{tr}(Y P) -\lambda s.
\)
where $Z \succeq 0$ is a four-block positive semi-definite matrix, $u\in \mathbb{R}^p$, $v\in \mathbb{R}^p$, $\lambda\ge 0$, lastly, $Y\in \mathbb{R}^{p\times p}$, except $Y_{i,j}\ge 0$ if $(i\to j)\in E_{\bar G}$. Clearly, the Lagrangian dual $\inf_{P,s} \mathcal L(\cdot)$ would be $-\infty$, unless:
\(
& -2 Z_{12}^{\rm T} + 1_m u^{\rm T} + v \pi_{\beta}^{\rm T}- Y=0,\\
& 1-\text{tr}(Z) -\lambda =0,
\)
which are equivalent to dual feasibility conditions:
\(
&Z_{12}(j,i)  \le  \frac{u_j + v_i \pi_\beta(j)}{2}  \text{ if } (i\to j)\in E_{\bar G}, \\
& \text{tr}(Z)\le 1, \;  Z\succeq 0
\)
for the dual problem:
\(
\sup_{Z, u,v}2  \text{tr} \big\{ Z^{\rm T}_{12} (1_m \pi^{\rm T}_\beta) \big\} - u^{\rm T}1_m -  \pi^{\rm T}_{\beta}v.
\)
We parameterize $Z= \tilde Z \tilde Z^{\rm T}$, with $\tilde Z\in\mathbb{R}^{p\times p}$, and use log-barrier to enforce inequalities, then use gradient ascent algorithm \{via the \texttt{JAX} package \citep{jax2018github}\} to find out $\hat Z,\hat u, \hat v$. Then using complementary slackness condition $s \cdot \text{tr}(Z_{11}+ Z_{22}) + 2\text{tr}\{Z_{12}^{\rm T}(P- 1_m \pi^{\rm T}_\beta)\}=0$ and primal optimal condition $s= \|P- 1_m \pi^{\rm T}_\beta\|_2$, we can find out the value of $\hat P$. We provide  numerical illustration in the Supplementary Materials S.1.

\bibliography{ref_fixed.bib}

\begin{thebibliography}{}

\bibitem[\protect\citeauthoryear{Betancourt, Byrne, Livingstone, and
  Girolami}{Betancourt et~al.}{2017}]{10.3150/16-BEJ810}
Betancourt, M., S.~Byrne, S.~Livingstone, and M.~Girolami (2017).
\newblock {The Geometric Foundations of Hamiltonian Monte Carlo}.
\newblock {\em Bernoulli\/}~{\em 23\/}(4A), 2257 -- 2298.

\bibitem[\protect\citeauthoryear{Bierkens}{Bierkens}{2016}]{bierkens2016non}
Bierkens, J. (2016).
\newblock {Non-Reversible Metropolis-Hastings}.
\newblock {\em Statistics and Computing\/}~{\em 26\/}(6), 1213--1228.

\bibitem[\protect\citeauthoryear{Bierkens, Fearnhead, and Roberts}{Bierkens
  et~al.}{2019}]{bierkens2019zig}
Bierkens, J., P.~Fearnhead, and G.~Roberts (2019).
\newblock {The Zig-Zag Process and Super-Efficient Sampling for Bayesian
  Analysis of Big Data}.
\newblock {\em The Annals of Statistics\/}~{\em 47\/}(3), 1288 -- 1320.

\bibitem[\protect\citeauthoryear{Blei, Kucukelbir, and McAuliffe}{Blei
  et~al.}{2017}]{blei2017variational}
Blei, D.~M., A.~Kucukelbir, and J.~D. McAuliffe (2017).
\newblock {Variational Inference: A Review for Statisticians}.
\newblock {\em Journal of the American Statistical Association\/}~{\em
  112\/}(518), 859--877.

\bibitem[\protect\citeauthoryear{Boyd, Diaconis, and Xiao}{Boyd
  et~al.}{2004}]{boyd2004fastest}
Boyd, S., P.~Diaconis, and L.~Xiao (2004).
\newblock {Fastest Mixing Markov Chain on a Graph}.
\newblock {\em SIAM Review\/}~{\em 46\/}(4), 667--689.

\bibitem[\protect\citeauthoryear{Bradbury, Frostig, Hawkins, Johnson, Leary,
  Maclaurin, Necula, Paszke, Vander{P}las, Wanderman-{M}ilne, and
  Zhang}{Bradbury et~al.}{2018}]{jax2018github}
Bradbury, J., R.~Frostig, P.~Hawkins, M.~J. Johnson, C.~Leary, D.~Maclaurin,
  G.~Necula, A.~Paszke, J.~Vander{P}las, S.~Wanderman-{M}ilne, and Q.~Zhang
  (2018).
\newblock {JAX}: Composable transformations of {P}ython+{N}um{P}y programs.

\bibitem[\protect\citeauthoryear{Bubeck}{Bubeck}{2015}]{Bubeck_2015}
Bubeck, S. (2015, November).
\newblock {Convex Optimization: Algorithms and Complexity}.
\newblock {\em Foundations and Trends® in Machine Learning\/}~{\em
  8\/}(3–4), 231–357.

\bibitem[\protect\citeauthoryear{Campbell and Li}{Campbell and
  Li}{2019}]{campbell2019universal}
Campbell, T. and X.~Li (2019).
\newblock {Universal Boosting Variational Inference}.
\newblock {\em Advances in Neural Information Processing Systems\/}~{\em 32}.

\bibitem[\protect\citeauthoryear{Chib and Carlin}{Chib and
  Carlin}{1999}]{chib1999mcmc}
Chib, S. and B.~P. Carlin (1999).
\newblock {On MCMC Sampling in Hierarchical Longitudinal Models}.
\newblock {\em Statistics and Computing\/}~{\em 9\/}(1), 17--26.

\bibitem[\protect\citeauthoryear{Costa and Dufour}{Costa and
  Dufour}{2008}]{costa2008stability}
Costa, O.~L. and F.~Dufour (2008).
\newblock {Stability and Ergodicity of Piecewise Deterministic Markov
  Processes}.
\newblock {\em SIAM Journal on Control and Optimization\/}~{\em 47\/}(2),
  1053--1077.

\bibitem[\protect\citeauthoryear{Dalalyan}{Dalalyan}{2017}]{dalalyan2017theoretical}
Dalalyan, A.~S. (2017).
\newblock {Theoretical Guarantees for Approximate Sampling From Smooth and
  Log-Concave Densities}.
\newblock {\em Journal of the Royal Statistical Society Series B: Statistical
  Methodology\/}~{\em 79\/}(3), 651--676.

\bibitem[\protect\citeauthoryear{Duan, Young, Nishimura, and Dunson}{Duan
  et~al.}{2020}]{duan2020bayesian}
Duan, L.~L., A.~L. Young, A.~Nishimura, and D.~B. Dunson (2020).
\newblock {Bayesian Constraint Relaxation}.
\newblock {\em Biometrika\/}~{\em 107\/}(1), 191--204.

\bibitem[\protect\citeauthoryear{Dupont, Doucet, and Teh}{Dupont
  et~al.}{2019}]{dupont2019augmented}
Dupont, E., A.~Doucet, and Y.~W. Teh (2019).
\newblock {Augmented Neural ODEs}.
\newblock In {\em {Advances in Neural Information Processing Systems}},
  Volume~32.

\bibitem[\protect\citeauthoryear{Durmus, Moulines, and Saksman}{Durmus
  et~al.}{2020}]{durmus2020irreducibility}
Durmus, A., {\'E}.~Moulines, and E.~Saksman (2020).
\newblock {Irreducibility and Geometric Ergodicity of Hamiltonian Monte Carlo}.

\bibitem[\protect\citeauthoryear{Durmus, Roberts, Vilmart, and
  Zygalakis}{Durmus et~al.}{2017}]{durmus2017fast}
Durmus, A., G.~O. Roberts, G.~Vilmart, and K.~C. Zygalakis (2017).
\newblock {Fast Langevin based algorithm for MCMC in high dimensions}.
\newblock {\em The Annals of Applied Probability\/}~{\em 27\/}(4), 2195--2237.

\bibitem[\protect\citeauthoryear{Fearnhead, Bierkens, Pollock, and
  Roberts}{Fearnhead et~al.}{2018}]{fearnhead2018piecewise}
Fearnhead, P., J.~Bierkens, M.~Pollock, and G.~O. Roberts (2018).
\newblock {Piecewise Deterministic Markov Processes for Continuous-Time Monte
  Carlo}.
\newblock {\em Statistical Science\/}~{\em 33\/}(3), 386--412.

\bibitem[\protect\citeauthoryear{Gabri{\'e}, Rotskoff, and
  Vanden-Eijnden}{Gabri{\'e} et~al.}{2022}]{gabrie2022adaptive}
Gabri{\'e}, M., G.~M. Rotskoff, and E.~Vanden-Eijnden (2022).
\newblock {Adaptive Monte Carlo Augmented with Normalizing Flows}.
\newblock {\em Proceedings of the National Academy of Sciences\/}~{\em
  119\/}(10), e2109420119.

\bibitem[\protect\citeauthoryear{Gelfand}{Gelfand}{2000}]{gelfand2000gibbs}
Gelfand, A.~E. (2000).
\newblock {Gibbs Sampling}.
\newblock {\em Journal of the American Statistical Association\/}~{\em
  95\/}(452), 1300--1304.

\bibitem[\protect\citeauthoryear{Gelfand, Smith, and Lee}{Gelfand
  et~al.}{1992}]{gelfand1992bayesian}
Gelfand, A.~E., A.~F. Smith, and T.-M. Lee (1992).
\newblock {Bayesian Analysis of Constrained Parameter and Truncated Data
  Problems Using Gibbs Sampling}.
\newblock {\em Journal of the American Statistical Association\/}~{\em
  87\/}(418), 523--532.

\bibitem[\protect\citeauthoryear{Gelman}{Gelman}{2004}]{gelman2004parameterization}
Gelman, A. (2004).
\newblock {Parameterization and Bayesian Modeling}.
\newblock {\em Journal of the American Statistical Association\/}~{\em
  99\/}(466), 537--545.

\bibitem[\protect\citeauthoryear{Gelman, Gilks, and Roberts}{Gelman
  et~al.}{1997}]{Gelman_Gilks_Roberts_1997}
Gelman, A., W.~R. Gilks, and G.~O. Roberts (1997).
\newblock {Weak Convergence and Optimal Scaling of Random Walk Metropolis
  Algorithms}.
\newblock {\em The Annals of Applied Probability\/}~{\em 7\/}(1), 110–120.

\bibitem[\protect\citeauthoryear{Giordano, Broderick, and Jordan}{Giordano
  et~al.}{2018}]{giordano2018covariances}
Giordano, R., T.~Broderick, and M.~I. Jordan (2018).
\newblock {Covariances, Robustness and Variational Bayes}.
\newblock {\em Journal of Machine Learning Research\/}~{\em 19\/}(51).

\bibitem[\protect\citeauthoryear{Girolami and Calderhead}{Girolami and
  Calderhead}{2011}]{girolami2011riemann}
Girolami, M. and B.~Calderhead (2011).
\newblock {Riemann Manifold Langevin and Hamiltonian Monte Carlo Methods}.
\newblock {\em Journal of the Royal Statistical Society: Series B (Statistical
  Methodology)\/}~{\em 73\/}(2), 123--214.

\bibitem[\protect\citeauthoryear{Haario, Saksman, and Tamminen}{Haario
  et~al.}{1999}]{haario1999adaptive}
Haario, H., E.~Saksman, and J.~Tamminen (1999).
\newblock {Adaptive Proposal Distribution for Random Walk Metropolis
  Algorithm}.
\newblock {\em Computational Statistics\/}~{\em 14}, 375--395.

\bibitem[\protect\citeauthoryear{Hoffman, Sountsov, Dillon, Langmore, Tran, and
  Vasudevan}{Hoffman et~al.}{2018}]{hoffman2019neutra}
Hoffman, M., P.~Sountsov, J.~V. Dillon, I.~Langmore, D.~Tran, and S.~Vasudevan
  (2018).
\newblock {Neutra-Lizing Bad Geometry in Hamiltonian Monte Carlo Using Neural
  Transport}.
\newblock In {\em {Symposium on Advances in Approximate Bayesian Inference}},
  pp.\  1--6.

\bibitem[\protect\citeauthoryear{Johndrow, Orenstein, and
  Bhattacharya}{Johndrow et~al.}{2020}]{johndrow2020scalable}
Johndrow, J., P.~Orenstein, and A.~Bhattacharya (2020).
\newblock {Scalable Approximate MCMC Algorithms for the Horseshoe Prior}.
\newblock {\em Journal of Machine Learning Research\/}~{\em 21\/}(73).

\bibitem[\protect\citeauthoryear{Johndrow, Smith, Pillai, and Dunson}{Johndrow
  et~al.}{2019}]{johndrow2019mcmc}
Johndrow, J.~E., A.~Smith, N.~Pillai, and D.~B. Dunson (2019).
\newblock {MCMC for Imbalanced Categorical Data}.
\newblock {\em Journal of the American Statistical Association\/}~{\em
  114\/}(527), 1394--1403.

\bibitem[\protect\citeauthoryear{Jones, Roberts, and Rosenthal}{Jones
  et~al.}{2014}]{jones2014convergence}
Jones, G.~L., G.~O. Roberts, and J.~S. Rosenthal (2014).
\newblock {Convergence of Conditional Metropolis-Hastings Samplers}.
\newblock {\em Advances in Applied Probability\/}~{\em 46\/}(2), 422--445.

\bibitem[\protect\citeauthoryear{Kong and Chaudhuri}{Kong and
  Chaudhuri}{2020}]{kong2020expressive}
Kong, Z. and K.~Chaudhuri (2020).
\newblock {The Expressive Power of a Class of Normalizing Flow Models}.
\newblock In {\em {International Conference on Artificial Intelligence and
  Statistics}}, pp.\  3599--3609. PMLR.

\bibitem[\protect\citeauthoryear{Kruskal}{Kruskal}{1956}]{kruskal1956shortest}
Kruskal, J.~B. (1956).
\newblock {On the Shortest Spanning Subtree of a Graph and the Traveling
  Salesman Problem}.
\newblock {\em Proceedings of the American Mathematical Society\/}~{\em
  7\/}(1), 48--50.

\bibitem[\protect\citeauthoryear{Lov{\'a}sz and Simonovits}{Lov{\'a}sz and
  Simonovits}{1992}]{lovasz1992randomized}
Lov{\'a}sz, L. and M.~Simonovits (1992).
\newblock {On the Randomized Complexity of Volume and Diameter}.
\newblock In {\em {Proceedings., 33rd Annual Symposium on Foundations of
  Computer Science}}, pp.\  482--492. IEEE Computer Society.

\bibitem[\protect\citeauthoryear{Lu and Lu}{Lu and Lu}{2020}]{lu2020universal}
Lu, Y. and J.~Lu (2020).
\newblock {A Universal Approximation Theorem of Deep Neural Networks for
  Expressing Probability Distributions}.
\newblock {\em Advances in Neural Information Processing Systems\/}~{\em 33},
  3094--3105.

\bibitem[\protect\citeauthoryear{Ma, Chatterji, Cheng, Flammarion, Bartlett,
  and Jordan}{Ma et~al.}{2021}]{10.3150/20-BEJ1297}
Ma, Y.-A., N.~S. Chatterji, X.~Cheng, N.~Flammarion, P.~L. Bartlett, and M.~I.
  Jordan (2021).
\newblock {Is There an Analog of Nesterov Acceleration for Gradient-Based
  MCMC?}
\newblock {\em Bernoulli\/}~{\em 27\/}(3), 1942 -- 1992.

\bibitem[\protect\citeauthoryear{Miller, Foti, and Adams}{Miller
  et~al.}{2017}]{miller2017variational}
Miller, A.~C., N.~J. Foti, and R.~P. Adams (2017).
\newblock {Variational Boosting: Iteratively Refining Posterior
  Approximations}.
\newblock In {\em {International Conference on Machine Learning}}, pp.\
  2420--2429. PMLR.

\bibitem[\protect\citeauthoryear{Murota}{Murota}{1998}]{murota1998discrete}
Murota, K. (1998).
\newblock {Discrete Convex Analysis}.
\newblock {\em Mathematical Programming\/}~{\em 83\/}(1-3), 313--371.

\bibitem[\protect\citeauthoryear{Natarovskii, Rudolf, and Sprungk}{Natarovskii
  et~al.}{2021}]{natarovskii2021geometric}
Natarovskii, V., D.~Rudolf, and B.~Sprungk (2021).
\newblock {Geometric Convergence of Elliptical Slice Sampling}.
\newblock In {\em {International Conference on Machine Learning}}, pp.\
  7969--7978. PMLR.

\bibitem[\protect\citeauthoryear{Neal}{Neal}{2003}]{neal2003slice}
Neal, R.~M. (2003).
\newblock {Slice Sampling}.
\newblock {\em The Annals of Statistics\/}~{\em 31\/}(3), 705--767.

\bibitem[\protect\citeauthoryear{Nishimura, Dunson, and Lu}{Nishimura
  et~al.}{2020}]{nishimura2020discontinuous}
Nishimura, A., D.~B. Dunson, and J.~Lu (2020).
\newblock {Discontinuous Hamiltonian Monte Carlo for Discrete Parameters and
  Discontinuous Likelihoods}.
\newblock {\em Biometrika\/}~{\em 107\/}(2), 365--380.

\bibitem[\protect\citeauthoryear{Pakman and Paninski}{Pakman and
  Paninski}{2013}]{pakman2013auxiliary}
Pakman, A. and L.~Paninski (2013).
\newblock {Auxiliary-Variable Exact Hamiltonian Monte Carlo Samplers for Binary
  Distributions}.
\newblock In {\em Advances in Neural Information Processing Systems},
  Volume~26.

\bibitem[\protect\citeauthoryear{Papamakarios, Nalisnick, Rezende, Mohamed, and
  Lakshminarayanan}{Papamakarios et~al.}{2021}]{papamakarios2021normalizing}
Papamakarios, G., E.~Nalisnick, D.~J. Rezende, S.~Mohamed, and
  B.~Lakshminarayanan (2021).
\newblock {Normalizing Flows for Probabilistic Modeling and Inference}.
\newblock {\em The Journal of Machine Learning Research\/}~{\em 22\/}(1),
  2617--2680.

\bibitem[\protect\citeauthoryear{Papaspiliopoulos, Roberts, and
  Sk{\"o}ld}{Papaspiliopoulos et~al.}{2007}]{papaspiliopoulos2007general}
Papaspiliopoulos, O., G.~O. Roberts, and M.~Sk{\"o}ld (2007).
\newblock {A General Framework for the Parametrization of Hierarchical Models}.
\newblock {\em Statistical Science\/}, 59--73.

\bibitem[\protect\citeauthoryear{Pasarica and Gelman}{Pasarica and
  Gelman}{2010}]{pasarica2010adaptively}
Pasarica, C. and A.~Gelman (2010).
\newblock {Adaptively Scaling the Metropolis Algorithm Using Expected Squared
  Jumped Distance}.
\newblock {\em Statistica Sinica\/}, 343--364.

\bibitem[\protect\citeauthoryear{Phan, Pradhan, and Jankowiak}{Phan
  et~al.}{2019}]{phan2019composable}
Phan, D., N.~Pradhan, and M.~Jankowiak (2019).
\newblock {Composable Effects for Flexible and Accelerated Probabilistic
  Programming in NumPyro}.
\newblock In {\em {Program Transformations for ML Workshop at NeurIPS 2019}}.

\bibitem[\protect\citeauthoryear{Pisier}{Pisier}{1999}]{pisier1999volume}
Pisier, G. (1999).
\newblock {\em {The Volume of Convex Bodies and Banach Space Geometry}},
  Volume~94.
\newblock Cambridge University Press.

\bibitem[\protect\citeauthoryear{Polson, Scott, and Windle}{Polson
  et~al.}{2013}]{polson2013bayesian}
Polson, N.~G., J.~G. Scott, and J.~Windle (2013).
\newblock {Bayesian Inference for Logistic Models Using P{\'o}lya--Gamma Latent
  Variables}.
\newblock {\em Journal of the American Statistical Association\/}~{\em
  108\/}(504), 1339--1349.

\bibitem[\protect\citeauthoryear{Pompe, Holmes, and Łatuszyński}{Pompe
  et~al.}{2020}]{pompe2020multimode}
Pompe, E., C.~Holmes, and K.~Łatuszyński (2020).
\newblock {A Framework for Adaptive MCMC Targeting Multimodal Distributions}.
\newblock {\em The Annals of Statistics\/}~{\em 48\/}(5), 2930 -- 2952.

\bibitem[\protect\citeauthoryear{Presman and Xu}{Presman and
  Xu}{2023}]{presman2023distance}
Presman, R. and J.~Xu (2023).
\newblock {Distance-to-Set Priors and Constrained Bayesian Inference}.
\newblock In {\em {International Conference on Artificial Intelligence and
  Statistics}}, pp.\  2310--2326. PMLR.

\bibitem[\protect\citeauthoryear{Prim}{Prim}{1957}]{prim1957shortest}
Prim, R.~C. (1957).
\newblock {Shortest Connection Networks and Some Generalizations}.
\newblock {\em The Bell System Technical Journal\/}~{\em 36\/}(6), 1389--1401.

\bibitem[\protect\citeauthoryear{Robert, Elvira, Tawn, and Wu}{Robert
  et~al.}{2018}]{robert2018accelerating}
Robert, C.~P., V.~Elvira, N.~Tawn, and C.~Wu (2018).
\newblock {Accelerating MCMC Algorithms}.
\newblock {\em Wiley Interdisciplinary Reviews: Computational
  Statistics\/}~{\em 10\/}(5), e1435.

\bibitem[\protect\citeauthoryear{Roberts and Polson}{Roberts and
  Polson}{1994}]{roberts1994geometric}
Roberts, G.~O. and N.~G. Polson (1994).
\newblock {On the Geometric Convergence of the Gibbs Sampler}.
\newblock {\em Journal of the Royal Statistical Society: Series B
  (Methodological)\/}~{\em 56\/}(2), 377--384.

\bibitem[\protect\citeauthoryear{Roberts and Rosenthal}{Roberts and
  Rosenthal}{1998}]{roberts1998optimal}
Roberts, G.~O. and J.~S. Rosenthal (1998).
\newblock {Optimal Scaling of Discrete Approximations to Langevin Diffusions}.
\newblock {\em Journal of the Royal Statistical Society: Series B (Statistical
  Methodology)\/}~{\em 60\/}(1), 255--268.

\bibitem[\protect\citeauthoryear{Roberts and Rosenthal}{Roberts and
  Rosenthal}{1999}]{roberts1999convergence}
Roberts, G.~O. and J.~S. Rosenthal (1999).
\newblock {Convergence of Slice Sampler Markov Chains}.
\newblock {\em Journal of the Royal Statistical Society: Series B (Statistical
  Methodology)\/}~{\em 61\/}(3), 643--660.

\bibitem[\protect\citeauthoryear{Roberts and Rosenthal}{Roberts and
  Rosenthal}{2001}]{Roberts_Rosenthal_2001}
Roberts, G.~O. and J.~S. Rosenthal (2001).
\newblock {Optimal Scaling for Various Metropolis-Hastings Algorithms}.
\newblock {\em Statistical Science\/}~{\em 16\/}(4), 351–367.

\bibitem[\protect\citeauthoryear{Roberts and Rosenthal}{Roberts and
  Rosenthal}{2004}]{roberts2004general}
Roberts, G.~O. and J.~S. Rosenthal (2004).
\newblock {General State Space Markov Chains and MCMC Algorithms}.

\bibitem[\protect\citeauthoryear{Roberts and Rosenthal}{Roberts and
  Rosenthal}{2009}]{roberts2009examples}
Roberts, G.~O. and J.~S. Rosenthal (2009).
\newblock {Examples of Adaptive MCMC}.
\newblock {\em Journal of Computational and Graphical Statistics\/}~{\em
  18\/}(2), 349--367.

\bibitem[\protect\citeauthoryear{Roberts and Smith}{Roberts and
  Smith}{1994}]{roberts1994simple}
Roberts, G.~O. and A.~F. Smith (1994).
\newblock {Simple Conditions for the Convergence of the Gibbs Sampler and
  Metropolis-Hastings Algorithms}.
\newblock {\em Stochastic Processes and Their Applications\/}~{\em 49\/}(2),
  207--216.

\bibitem[\protect\citeauthoryear{Roberts and Tweedie}{Roberts and
  Tweedie}{1996}]{roberts1996exponential}
Roberts, G.~O. and R.~L. Tweedie (1996).
\newblock {Exponential Convergence of Langevin Distributions and Their Discrete
  Approximations}.
\newblock {\em Bernoulli\/}, 341--363.

\bibitem[\protect\citeauthoryear{Rue, Martino, and Chopin}{Rue
  et~al.}{2009}]{rue2009approximate}
Rue, H., S.~Martino, and N.~Chopin (2009).
\newblock {Approximate Bayesian Inference for Latent Gaussian Models by Using
  Integrated Nested Laplace Approximations}.
\newblock {\em Journal of the Royal Statistical Society Series B: Statistical
  Methodology\/}~{\em 71\/}(2), 319--392.

\bibitem[\protect\citeauthoryear{Sohl-Dickstein, Mudigonda, and
  DeWeese}{Sohl-Dickstein et~al.}{2014}]{sohl2014hamiltonian}
Sohl-Dickstein, J., M.~Mudigonda, and M.~DeWeese (2014).
\newblock {Hamiltonian Monte Carlo Without Detailed Balance}.
\newblock In {\em International Conference on Machine Learning}, pp.\
  719--726. PMLR.

\bibitem[\protect\citeauthoryear{Tang and Yang}{Tang and
  Yang}{2024}]{tang2024computational}
Tang, R. and Y.~Yang (2024).
\newblock {On the Computational Complexity of Metropolis-Adjusted Langevin
  Algorithms for Bayesian Posterior Sampling}.
\newblock {\em Journal of Machine Learning Research\/}~{\em 25\/}(157), 1--79.

\bibitem[\protect\citeauthoryear{Tanner and Wong}{Tanner and
  Wong}{1987}]{tanner1987calculation}
Tanner, M.~A. and W.~H. Wong (1987).
\newblock {The Calculation of Posterior Distributions by Data Augmentation}.
\newblock {\em Journal of the American Statistical Association\/}~{\em
  82\/}(398), 528--540.

\bibitem[\protect\citeauthoryear{Toth, Rezende, Jaegle, Racanière, Botev, and
  Higgins}{Toth et~al.}{2020}]{Toth2020Hamiltonian}
Toth, P., D.~J. Rezende, A.~Jaegle, S.~Racanière, A.~Botev, and I.~Higgins
  (2020).
\newblock {Hamiltonian Generative Networks}.
\newblock In {\em {International Conference on Learning Representations}}.

\bibitem[\protect\citeauthoryear{Vershynin}{Vershynin}{2015}]{vershynin2015estimation}
Vershynin, R. (2015).
\newblock {Estimation in High Dimensions: A Geometric Perspective}.
\newblock In {\em {Sampling Theory, a Renaissance: Compressive Sensing and
  Other Developments}}, pp.\  3--66. Springer.

\bibitem[\protect\citeauthoryear{Vershynin}{Vershynin}{2018}]{vershynin2018high}
Vershynin, R. (2018).
\newblock {\em {High-Dimensional Probability: An Introduction with Applications
  in Data Science}}, Volume~47.
\newblock Cambridge university press.

\bibitem[\protect\citeauthoryear{Vono, Paulin, and Doucet}{Vono
  et~al.}{2022}]{vono2022efficient}
Vono, M., D.~Paulin, and A.~Doucet (2022).
\newblock {Efficient MCMC Sampling with Dimension-Free Convergence Rate Using
  ADMM-type Splitting}.
\newblock {\em The Journal of Machine Learning Research\/}~{\em 23\/}(1),
  1100--1168.

\bibitem[\protect\citeauthoryear{Winter, Campbell, Lin, Srivastava, and
  Dunson}{Winter et~al.}{2024}]{winter2024emerging}
Winter, S., T.~Campbell, L.~Lin, S.~Srivastava, and D.~B. Dunson (2024).
\newblock {Emerging Directions in Bayesian Computation}.
\newblock {\em Statistical Science\/}~{\em 39\/}(1), 62--89.

\bibitem[\protect\citeauthoryear{Yi, Liu, Liu, and Zhou}{Yi
  et~al.}{2023}]{yi2023global}
Yi, S.-Y., Z.~Liu, M.-Q. Liu, and Y.-D. Zhou (2023).
\newblock {Global Likelihood Sampler for Multimodal Distributions}.
\newblock {\em Journal of Computational and Graphical Statistics\/}, 927--937.

\end{thebibliography}
\bibliographystyle{chicago}

\section*{Supplementary Materials}

\section*{S.1 Numerical illustration on different choices of graph}

\begin{figure}[H]
 \begin{subfigure}[t]{0.24\textwidth}
 \centering
      \begin{overpic}[width=1.0\linewidth]{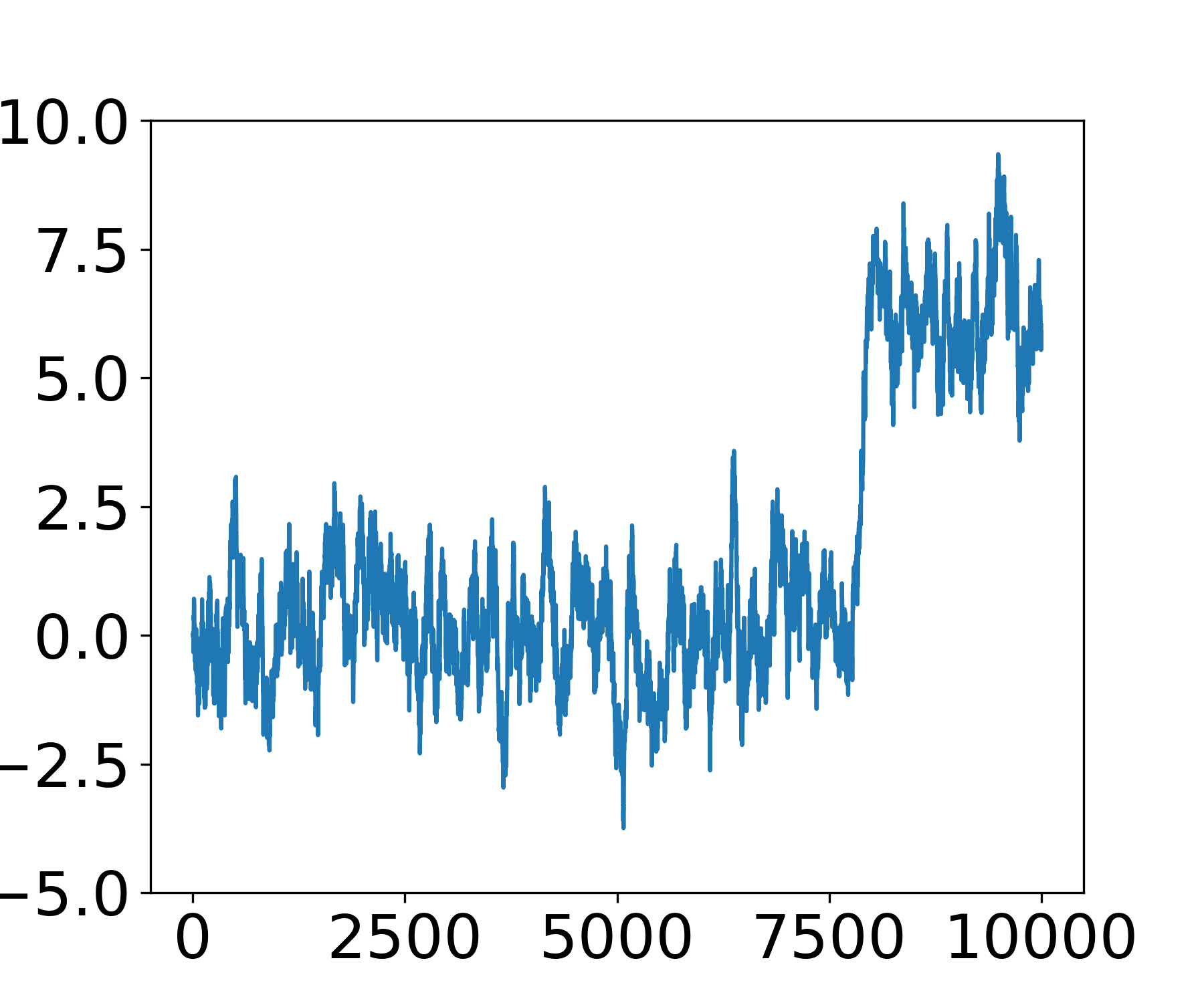}
    \end{overpic}
\caption{\footnotesize  Traceplot of $\theta_2$ using random-walk Metropolis.}
 \end{subfigure}\;
   \begin{subfigure}[t]{.24\textwidth}
 \centering
       \begin{overpic}[width=1.0\linewidth]{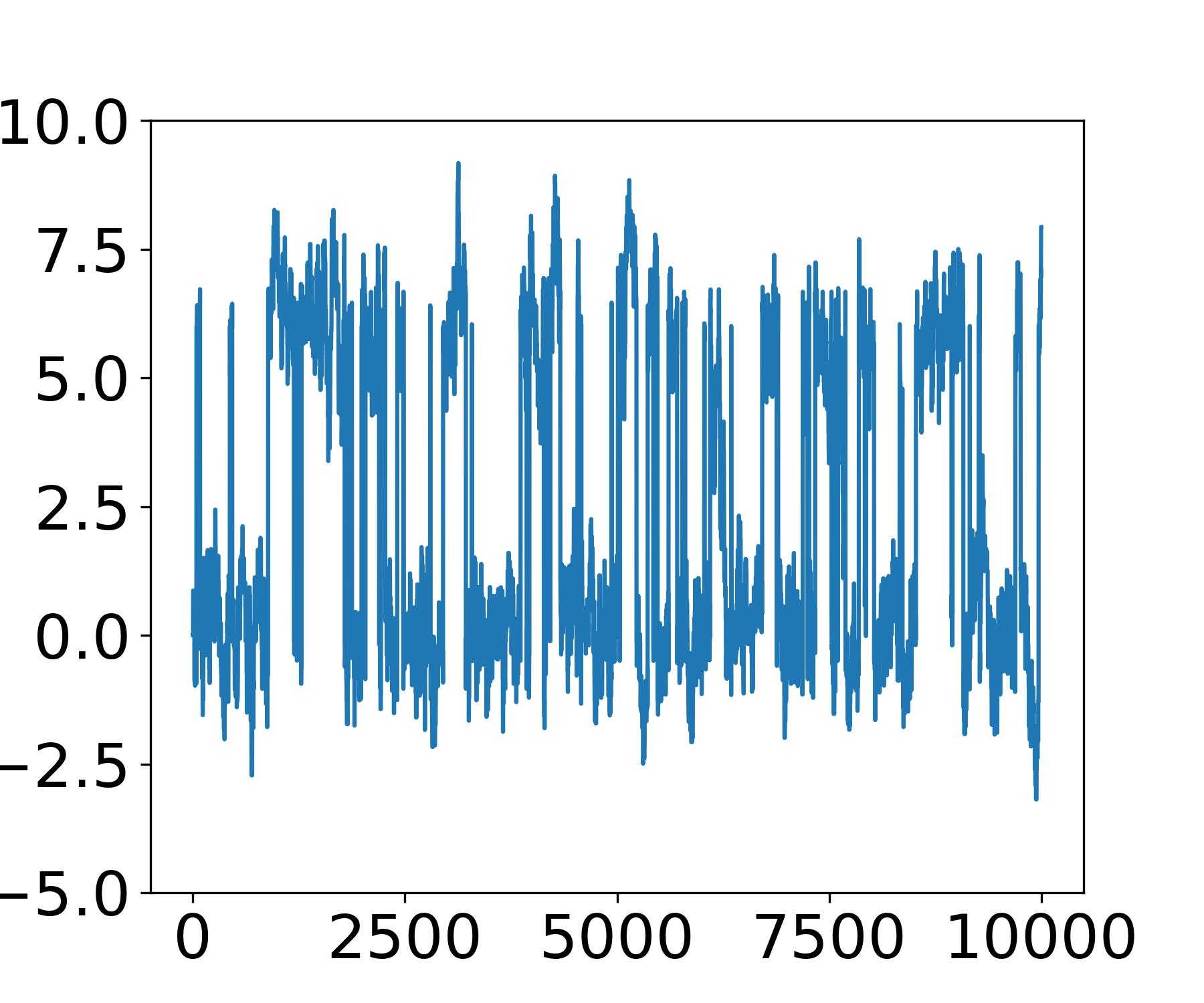}
    \end{overpic}
\caption{\footnotesize  Traceplot of $\theta_2$ using accelerated algorithm with spanning tree graph with $r=1$, $\kappa=1$.}
                       \end{subfigure}\;
 \begin{subfigure}[t]{0.24\textwidth}
 \centering
        \begin{overpic}[width=1.0\linewidth]{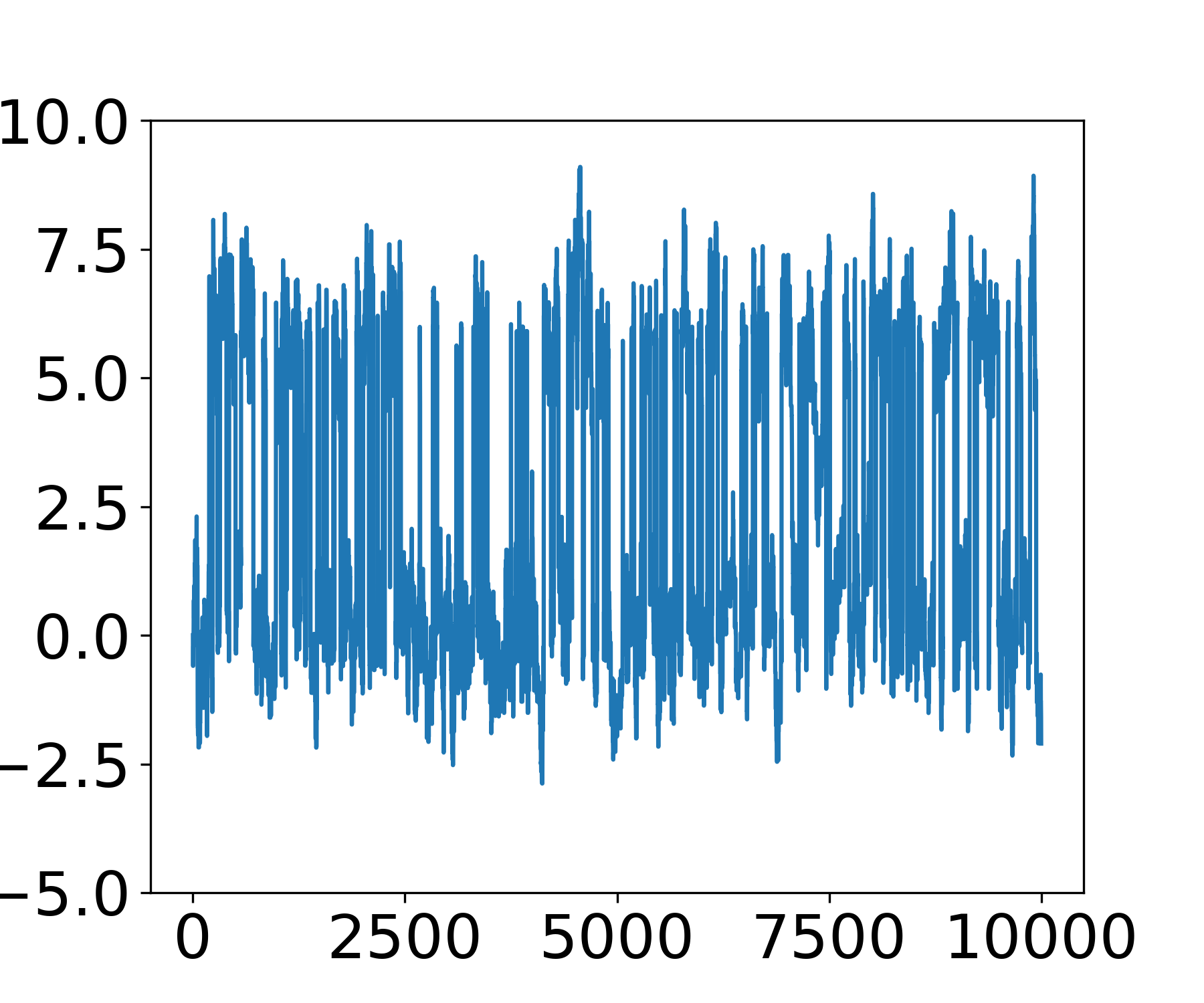}
    \end{overpic}
\caption{\footnotesize  Traceplot of $\theta_2$ using acceleration via spanning tree graph with greedily optimized $(r,\kappa)$.}
 \end{subfigure}\;
  \begin{subfigure}[t]{0.24\textwidth}
 \centering
         \begin{overpic}[width=1.0\linewidth]{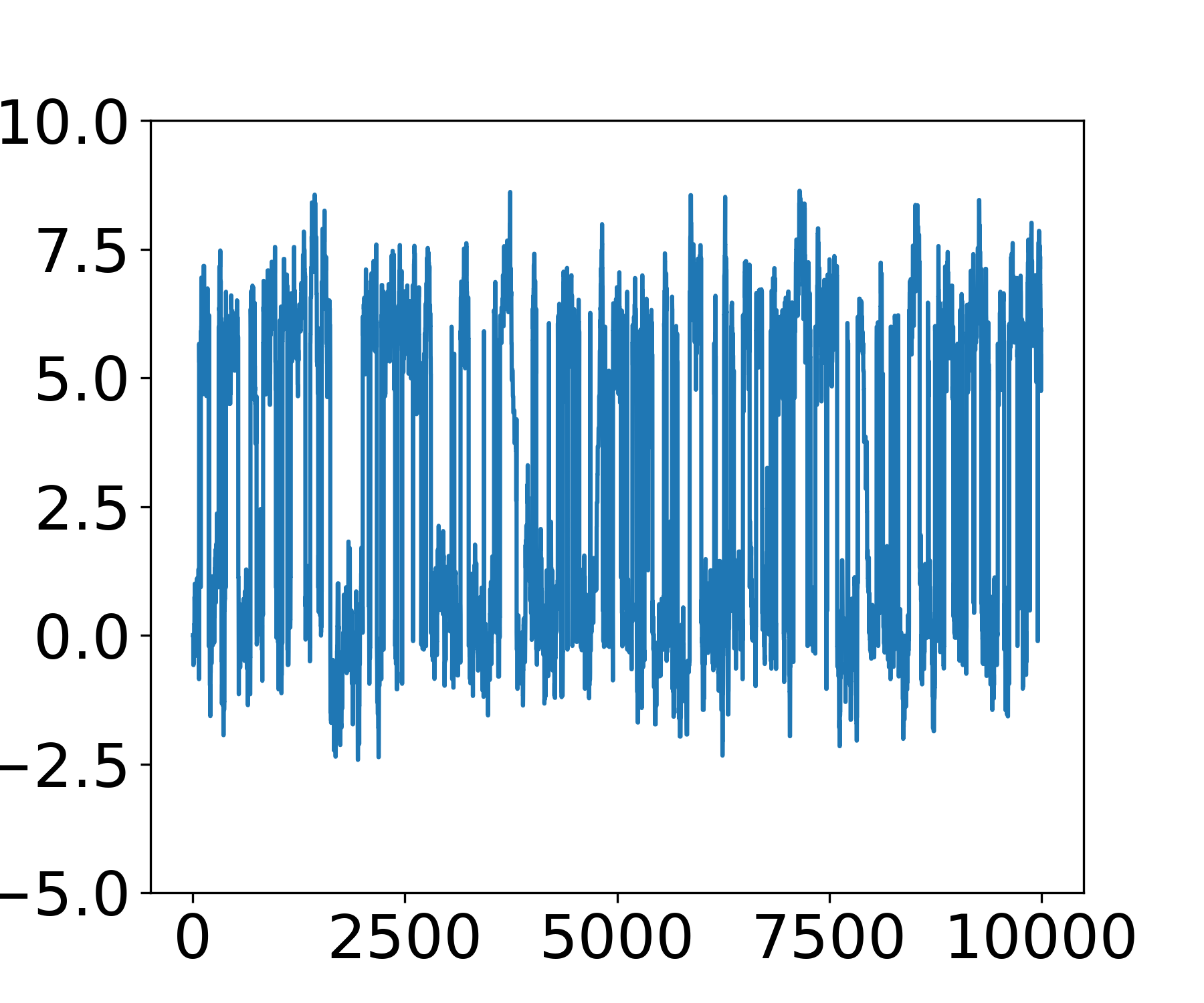}
    \end{overpic}
\caption{\footnotesize  Traceplot of $\theta_2$ using acceleration via  optimized random walk graph.}
 \end{subfigure}
  \caption{Comparing acceleration algorithms using different graphs, for sampling a two-component Gaussian mixture distribution. \label{fig:gmm_comp}}
 \end{figure}

For numerical illustration, we use the Gaussian mixture example we consider in the main text. In addition to (i) the baseline random walk Metropolis and (ii) the accelerated algorithm with spanning tree graph under default value $(r=1,\kappa=1)$, we experiment with (iii) the accelerated algorithm with greedy optimization on $(r,\kappa)$ to maximize the expected squared jumped distance (with $r=5$ and $\kappa=0.65$), and (iv) the accelerated algorithm using optimized random walk (with edges excluded if $\|\beta^i-\beta^j\|\le 0.5$). For (ii)(iii) and (iv), we use the same collection of $m=100$ samples, and we compare the mixing performance of those algorithm via the traceplot of $\theta_2$ in Figure \ref{fig:gmm_comp}. As can be seen, (iii) and (iv) further improve the mixing compared to (ii), although these two extensions are much more complicated.

\section*{S.2 Estimate on the vanishing rate of acceptance probability of Gaussian random-walk Metropolis algorithm}

It has been shown 
  \citep{Gelman_Gilks_Roberts_1997,Roberts_Rosenthal_2001} that for Gaussian random-walk Metropolis algorithm with target distribution consisting of $p$ independent components,  one may use a Gaussian proposal with standard deviation at $ c p^{-1/2}$, so that the acceptance rate could stay above zero and converge to $2\Phi(- \tilde m   c )$ as $p\to \infty$, with some $\tilde m>0$ depending on the target distribution and $\Phi$ the standard Gaussian cumulative distribution function. 
  
  To roughly estimate the vanishing speed of the acceptance rate under a fixed step size, we can replace $ c$ by $ c p^{1/2}$ and obtain $2\Phi(- \tilde m c  p^{1/2} )$. For $x>0$ and $t>x$,  $$\Phi(-x) = (2\pi)^{-1/2} \int_x^\infty \exp(-t^2/2)\textup{d}t 
 \le (2\pi)^{-1/2} \int_x^\infty (t/x) \exp(-t^2/2)\textup{d}t
 = (2\pi)^{-1/2} \exp(-x^2/2)/x.$$ Plugging $x=mcp^{1/2}$ yields   $O(p^{-1/2} \exp\{- \tilde c p)\}$ for some constant $\tilde c>0$. Omitting the dominated $p^{-1/2}$ leads to the $O(\exp\{- \tilde c p)\}$ rate.

\section*{S.3 Numerical results on the change of acceptance probability}
To show empirical evidence that the acceptance rate remains positive and away from zero in practice, we adopt the same latent Gaussian model used in the application, but now fit the model to simulated data of different sample size $n\in \{100,500,1000,2000\}$. We use $\tau = 1$, $h = 0.25$, and $r = 2$ during data simulation. Since each data point $y_i$ is associated with a latent $z_i$, the effective dimension of variables to sample is $p=(n+3)$. We run the Gibbs sampling algorithm (as the baseline algorithm described  in the main text) for $2000$ iterations, with first $400$ discarded as burn-in, then take a subset of size $m$ as the approximate samples. In each experiment, we run the accelerated algorithm for $2000$ iterations with $w=0.5$ and $r=3$, and report the empirical acceptance rate as the number of accepted graph jump steps divided by $1000$, as equal to $2000 \times 0.5$.

\begin{figure}[H]\centering
 \begin{subfigure}[t]{0.45\textwidth}
      \begin{overpic}[width=1.0\linewidth]{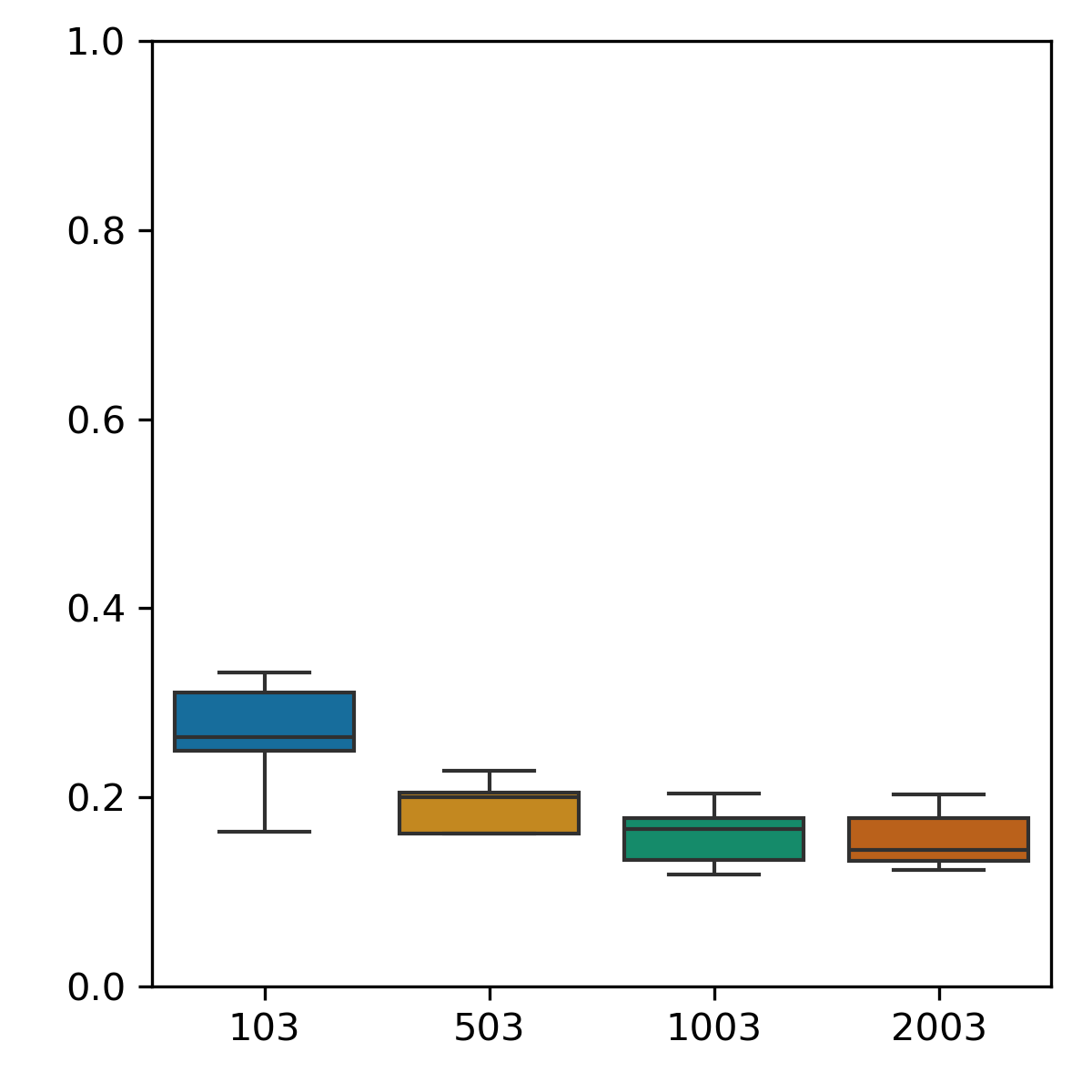}
    \put(50, -4){\scriptsize Dimension}
    \put(-3,30){\rotatebox{90}{\scriptsize M-H Acceptance Rate}}
    \end{overpic}
\caption{\footnotesize  Acceptance rates versus different dimensions $p$, under fixed $m=1600$.}
 \end{subfigure}\qquad
   \begin{subfigure}[t]{.45\textwidth}
 \centering
       \begin{overpic}[width=1.0\linewidth]{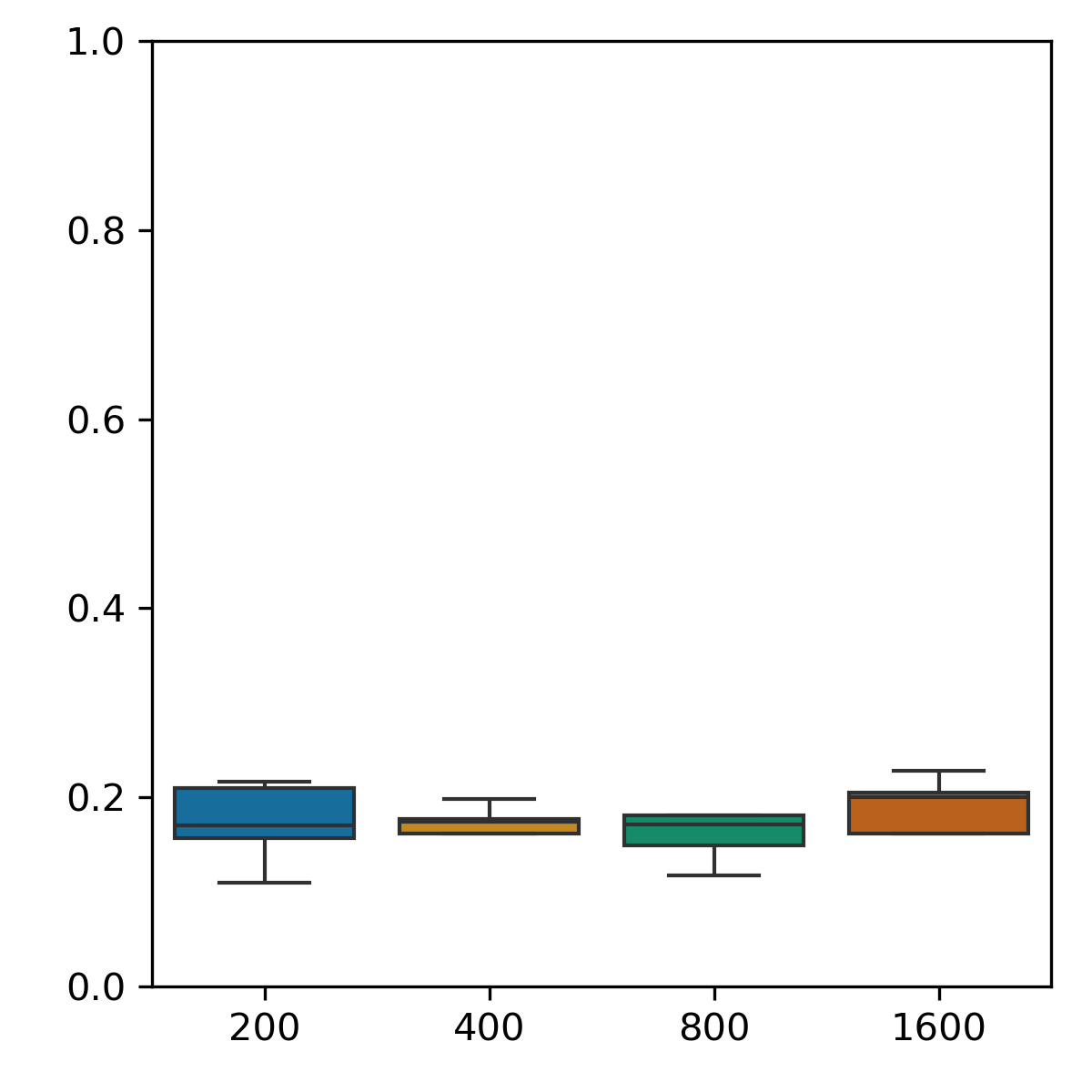}
    \put(30, -4){\scriptsize Approximate Sample Size}
    \put(-3,30){\rotatebox{90}{\scriptsize M-H Acceptance Rate}}
    \end{overpic}
\caption{\footnotesize  Acceptance rates versus different approximate sample sizes $m$, under fixed $p=503$.}
                       \end{subfigure}
  \caption{Numerical result of the Metropolis-Hastings acceptance rate in the graph jump step. The acceptance rates are calculated from simulated experiments of posterior sampling from a latent Gaussian model. \label{fig:accept_rate}}
 \end{figure}

We first conduct experiments under $m=1600$ and different $p \in \{103,503,1003,2003\}$. We repeat the experiments for $5$ times under each value of $p$. As shown in the boxplots in Figure \ref{fig:accept_rate}(a), the decrease of the Metropolis-Hastings acceptance rate over $p$ is slow: on average, the acceptance rate is around 26\% when $p=103$, and around 17\% when $p=2003$. Next, we conduct experiments under $p=503$ with different sizes of approximate samples $m\in\{200,400,800,1600\}$. As shown in the boxplots in Figure \ref{fig:accept_rate}(b), the average acceptance rates show no clear difference under different $m$  in the range.

\end{document}